# High-Crystalline-Fidelity Transparent Conductive Perovskites with Pronounced Chemical Disorder

*[1]Saeed S. I. Almishal, [2]Pat Kezer, [1]Yasuyuki Iwabuchi, [3]Jacob T. Sivak, [1]Sai Venkata Gayathri Ayyagari, [1]Saugata Sarker, [1]Matthew Furst, [5]Gerald Bejger, [1]Billy Yang, [1]Simon Gelin, [1]Nasim Alem, [1]Ismaila Dabo, [5]Christina M. Rost, [1,3,6]Susan B. Sinnott, [4]Vincent Crespi, [1]Venkatraman Gopalan, [7] Roman Engel-Herbert, [8] John T. Heron, *[1]Jon-Paul Maria

[1]*Department of Materials Science and Engineering, The Pennsylvania State University, University Park, PA 16802, USA*

[2]*Department of Electrical Engineering, University of Michigan, Ann Arbor, MI, 48109, USA*

[3]*Department of Chemistry, The Pennsylvania State University, University Park, PA 16802, USA*

[4]*Department of Physics, The Pennsylvania State University, University Park, PA 16802, USA*

[5]*Department of Materials Science and Engineering, Virginia Polytechnic Institute and State University, Blacksburg, VA 24061, USA*

[6]*Institute for Computational and Data Sciences, The Pennsylvania State University, University Park, PA 16802, USA*

[7]*Paul-Drude Institute for Solid State Electronics, Hausvogteiplatz 5-7, 10117 Berlin, Germany*

[8]*Department of Materials Science and Engineering, University of Michigan, Ann Arbor, MI, 48109, USA*

Corresponding authors: Saeed S. I. Almishal saeedsialmishal@gmail.com and Jon-Paul Maria jpm133@psu.edu

**Abstract**

This manuscript presents a working model linking chemical disorder and transport properties in correlated-electron perovskites with high-entropy formulations and a framework to actively design them. We demonstrate this new learning in epitaxial $Sr_x(Ti,Cr,Nb,Mo,W)O_3$ thin films that exhibit exceptional crystalline fidelity despite a diverse chemical formulation where most *B*-site species are highly misfit with respect to valence and radius. X-ray diffraction, X-ray photoelectron spectroscopy, and transmission electron microscopy confirm a unique combination of chemical disorder and structural perfection in thick epitaxial layers. This combination produces significant electron correlation, low electrical resistivity, and an optical transparency window that surpasses that of constituent end-members, with a flattened frequency- and temperature-dependent response. We address the computational challenges of modeling such systems and investigate short-range ordering using cluster expansion. These results showcase that unusual *d*-metal combinations access an expanded property design space that is predictable using end-member characteristics – though unavailable to them – thus offering performance advances in optical, spintronic, and quantum devices.

Strongly correlated materials host exotic phenomena induced by electron-electron interactions such as high-temperature superconductivity and quantum magnetism, possibly enriched by topological phenomena. These emergent properties could revolutionize future technologies such as quantum computation.[1–7] Correlated oxides in the A*B*$O_3$ perovskite structure attract significant attention because their structure and properties are tunable through cation substitution, and they can be integrated into functional heterostructures and superlattices.[1,8] $Sr_xBO_3$ cubic perovskites are particularly interesting for high electrical conductivity especially when early transition metal cations with n$d^1$ and n$d^2$ electronic configurations occupy the *B*-site.[1,2,9–14] Remarkably, the $Sr_xBO_3$ cubic perovskites remain paramagnetic metals (with no metal-to-insulator transitions near room temperature) with an optical transparency window that results from electron correlation. $Sr_xVO_3$, $Sr_xNbO_3$ and $Sr_xMoO_3$ are thus extensively studied as transparent conductors.[2,9,11–13] In contrast, $Sr_xCrO_3$ (with 3$d^3$ $Cr^{3+}$ on the *B*-site) and $Sr_xWO_3$ (whose *B*-site W cations have large spin-orbit coupling (SOC)) remain relatively unexplored due to their highly metastable nature in the cubic perovskite phase. In fact, all $Sr_xBO_3$ family members thermodynamically prefer insulating phases like scheelite and barite unless heroically engineered with significant vacancies on the *A*-site and grown under highly reducing conditions.[10,15,16]

High-entropy oxides (HEOs) can stabilize structure-formulation combinations with unusual coordination and valence[17,18], thus a possible new route to prepare early-transition-metal $Sr_xBO_3$ perovskites. HEOs maximize chemical configurational entropy on equivalent cation sublattice sites while maintaining positional order, which tends to thermodynamically favor homogeneous high-symmetry phases. Such entropy-favored phases can be further promoted by quenching energetic plasma adatoms on a comparatively cool substrate to capture a high effective-temperature structure and kinetically arrest further transitions to equilibrium.[19–21] As such, combining many-cation chemical formulations with far-from-equilibrium synthesis methods boosts both entropy and temperature, potentially realizing crystals that violate Muller and Roy's structure field maps while retaining high crystalline fidelity and significant chemical disorder (Taxonomy: Supplementary Note 1).[17,18,20,22,23] We therefore hypothesize that a $Sr_xBO_3$ formulation with many cations on the *B*-site can promote the high-symmetry cubic perovskite phase, solvating a diverse palette of early and heavy transition metals and unlocking novel correlation-disorder induced properties. This approach may enable transparent conductors with a more tunable UV-to-infrared optical response, heightened SOC with controllable resistivity, spin-orbit torque devices, and high-energy plasmonic systems.[7,11,24,25] To initiate learning in disordered correlated perovskite metals, we explore the relationships linking optical transparency and electrical conductivity. We first formulate design rules that identify $Sr_x$(Ti,Cr,Nb,Mo,W)$O_3$ as a candidate composition in this space, then use high kinetic energy synthesis to realize high-quality $Sr_x$(Ti,Cr,Nb,Mo,W)$O_3$ films that are practically metastable, though far from equilibrium under room conditions. In parallel, we develop a cluster expansion model to predict tendencies for short-range order in such materials as a function of temperature. Through this process we demonstrate an electrical conductivity-optical transparency combination that surpasses endmember possibilities.

## 1 Designing chemical disorder on the perovskite B-site for optical transparency and electrical conductivity

We begin with a computational exploration of $SrBO_3$ endmember band structures using density functional theory (DFT) to inform predictions of electronic structure and properties in many-component solid solutions. As an example, Figure 1(a) presents the $SrMoO_3$ band structure. Three $dt_{2g}$ bands originate from the 4*d* orbital manifold with a $EW_{Modt2g}$ bandwidth (where $EW_{Bdt2g}$ denotes $dt_{2g}$ energy bandwidth for cation B). These bands are partially filled by Mo *d*-electrons and energetically isolated from the $O_{2p}$ bands, forming a *buried energy gap,* $E_{O2p\text{-}Modt2g}$. Figure 1(b) shows a transmission (%) vs. energy schematic for a hypothetical conducting perovskite oxide. The transparency window is bounded by a reflection edge on the low-energy side and an absorption edge on the high-energy side. As we argue below, these physical properties are tied directly to the descriptor energies $EW_{Bdt2g}$ and $E_{O2p\text{-}Bt2g}$, respectively. The overall band structure of our perovskite endmembers share a similar form and exhibit descriptor energies that vary systematically with elemental periodic properties (Supplementary Note 2). In most general terms, a smaller $EW_{Bdt2g}$ implies smaller *d*-orbital overlap, stronger correlation and thus a larger electron effective mass. This effectively redshifts the plasma frequency $\omega_p$ and the associated IR reflection edge; Figures 1(c) and (d) show this periodic trend. Based on this criterion, $SrVO_3$ and $SrCrO_3$ have the strongest correlation and are predicted to enhance transparency in the low-energy visible and IR regions.

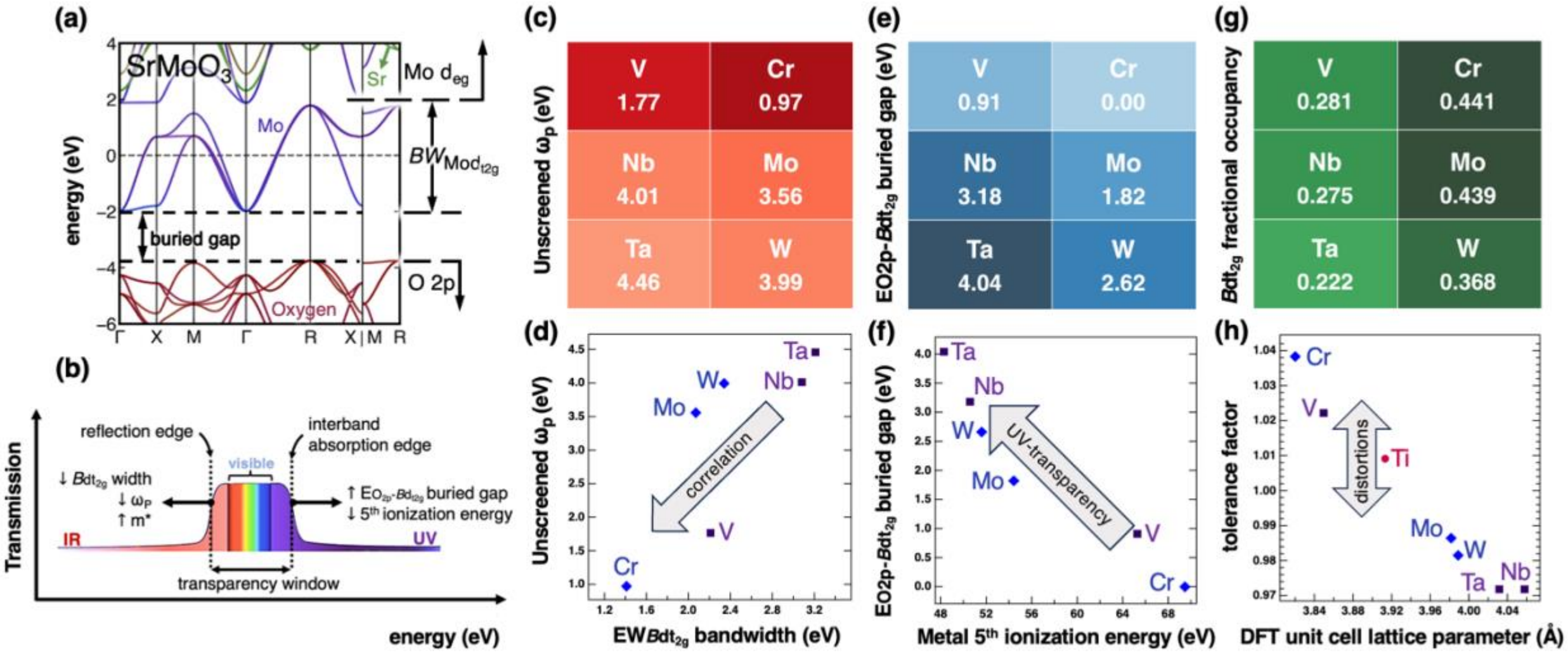

**Figure 1. B-cation design criteria.** (a) $SrMoO_3$ calculated band structure, (b) engineering the transparency window, (c) unscreened plasma frequency map for 5B and 6B metals, (d) the relation between bandwidth $W_{Bdt2g}$ and plasma frequency $\omega_p$: greater d-orbital overlap leads to lower electron-electron correlation, a larger $B_{t2g}$ and a lower energy plasma frequency (we rely on $W_{Bdt2g}$ and $\omega_p$ as correlation indicators rather than the typical renormalization factor $Z_k$) (e) map for buried energy gap $E_{O2p\text{-}t2g}$ defined as the gap between the highest energy lying $O_{2p}$ state and the lowest energy $dt_{2g}$ state,(f) the correlation between the buried gap and the 5[th] ionization energy, (g) $dt_{2g}$ fractional occupancy map and (h) tolerance factor vs the unit cell lattice parameters obtained from DFT relaxation. (c)-(h) are calculated from supercells with anti-ferromagnetic type-G (AFM-G) ordering, which better approximates strong correlation than do nonmagnetic (NM) unit cells (Supplementary Note 2).[26,27]

On the opposite end, the $E_{O2p\text{-}Bdt2g}$ energy gap can be related to the high energy absorption edge, which originates from interband transitions between occupied $O_{2p}$ bands and unoccupied $Bndt_{2g}$ states above $E_F$, where a larger $E_{O2p\text{-}Bdt2g}$ gap blue-shifts the transparency window into the UV spectrum. We note that prior authors suggested that this edge scales with orbital size or electronegativity difference between *B* and O.[9–11]This trend notably fails for W and Mo, for which electronegativity models predict UV transparency but experiment finds absorption. Our $E_{O2p\text{-}Bdt2g}$ descriptor, however, correctly assigns $SrNbO_3$ as more transparent than $SrMoO_3$ and $SrWO_3$, and importantly, offers some chemical intuition given its relationship to the *B*-site metal cation 5th ionization energy ($E_i^5$) which is shown in Figure 1(f). For the present series, $E_i^5$ values follow the hierarchy: Cr > V > Mo > W > Nb > Ta. Decreasing $E_i^5$ values indicate a willingness for $5^+$ oxidation states, which in turn depopulates the $dt_{2g}$ bands, opens the $O_{2p}$-$B_{dt2g}$ band energy separation and blue-shifts the high energy absorption edge. A correlation reduction, however, usually accompanies this blue shift which reduces slightly the near-IR transparency. With these criteria the broadest transparency window with high conductivity is expected for $SrNbO_3$.

$B_{dt2g}$ filling is the critical design parameter to optimize carrier density, the fractional filling associated with each *B*-site cation is listed in Figure 1(g). In addition to carrier density, band filling fraction can also increase correlation effects (captured when enforcing antiferromagnetic type-G ordering and is not observed in the nonmagnetic case, see Supplementary Note 1) and contribute to structural distortions from cubic perovskite by filling antibonding states as shown in Figure 1(h). $SrMoO_3$ with 44% $Mo4d_{t2g}$ predicted occupancy, has the lowest experimentally-reported electrical resistivity with the highest carrier concentration in this family.[13,16,26] Following these trends, we hypothesize that cubic perovskites with high Cr and W concentrations will exhibit high electrical conductivity because $t_{2g}$ filling will rival $SrMoO_3$. Additionally, Cr and W, as crystallochemical misfits, could lift centrosymmetry at short length scales intensifying electronic correlations.

With these theory trends in mind, we propose $Sr_x$(Ti,Cr,Nb,Mo,W)$O_3$ as the ideal host optimizing high conductivity, high transparency and high spin-orbit coupling. Ta and V are excluded due to tendencies for insulating behavior and toxicity respectively, while Ti is added to stabilize the perovskite structure. We note that Cr, Nb, Mo, and W are not perovskite *B*-site cations at ambient equilibrium, thus part two of our guiding hypothesis is entropic stabilization of the essential perovskite structure for reasons articulated earlier.

## 2 Enhancing the stability of disordered cubic perovskite phases beyond that of end-members through nonequilibrium thin film growth

We anticipate that establishing perovskite Sr(Ti,Cr,Nb,Mo,W)$O_3$ requires quenching high kinetic energy adatoms and substrate epitaxial constraints. In addition, engineering *A*-site vacancies to balance *B*-cations with > $4^+$ net valence is also necessary. We therefore first prepare bulk ceramics with 0% to 25% Sr deficiency (Supplementary Note 3). These ceramics, sintered at 1400°C, exhibit both scheelite and perovskite phases, but pulsed laser deposition at 850°C in 50 mTorr Ar produces epitaxial perovskite material with high crystalline fidelity on multiple substrates. Processing details needed to stabilize perovskite films are discussed in Methods and Supplementary Notes 4 and 12. Ultimately, ceramic targets with 5% A-site vacancies produce superior crystals and properties (Supplementary Note 4). We further investigate this composition, henceforth referred to as $Sr_{0.95}BO_3$.

Figure 2(a) depicts X-ray diffraction (XRD) scans for $Sr_{0.95}BO_3$ films grown on different technologically relevant substrates: (001) $(LaAlO_3)_{0.3}(Sr_2TaAlO_6)_{0.7}$ (LSAT), (001) $SrTiO_3$, (110) $DyScO_3$, (110) $GdScO_3$ and (001) $KTaO_3$ (Supplementary Note 5). The corresponding wide-angle 2θ−ω XRD scans in Figure S6 show that only perovskite phase peaks are present. The strong diffraction peaks with pronounced Pendellösung fringes in Figure 2(a) indicate abrupt interfaces and smooth film surfaces, also confirmed by atomic force microscopy in Figure S7. We choose LSAT substrates for thickness and electrical studies due to their insulating properties and ability to produce high-quality crystals. Figure 2(b) illustrates four $Sr_{0.95}BO_3$ films grown on LSAT with increasing thickness up to 54 nm. All films exhibit high crystalline quality with low mosaicity (rocking curve widths all below 0.05° using BBHD optics), despite substantial chemical disorder and that 4 *B*-site cations prefer different equilibrium structures (Supplementary Note 6 and Figure S9). This crystalline fidelity and epitaxial growth persists to 300 nm thickness with only partial broadening between 200 and 300 nm (Figure S10). This is significantly thicker than previously reported end-member thin films, which typically show full relaxation or noticeable degradation in crystalline quality beyond 60 nm[2,9,10]. The asymmetric $\bar{1}03$ diffraction peak reciprocal space map (RSM) for the 35 nm film is shown in Figure 2(c). From the RSM and out-of-plane symmetric scans, we calculate out-of-plane and in-plane lattice constants of 3.983 Å (± 005Å) and 3.963 Å (± 005Å) respectively, yielding a c/a ratio ~ 1.00. We calculate the film's intrinsic lattice parameter to be 3.967 Å, corresponding to a 2.5% lattice mismatch with the 3.87 Å LSAT substrate. Moreover, the RSM in Figure 2(c) suggests film relaxation, though it remains uncertain if this process is partial or complete.

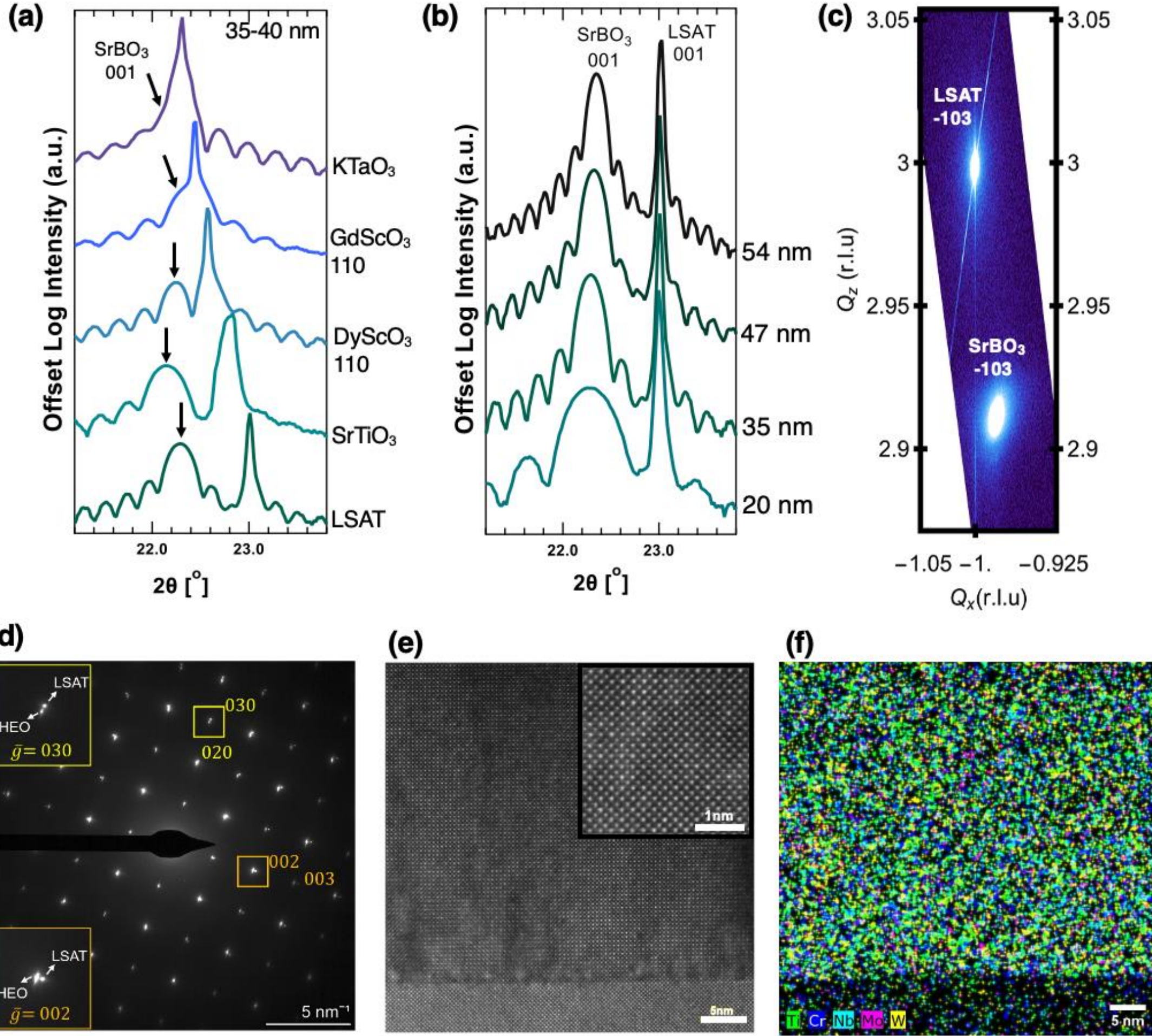

**Figure 2. $Sr_{0.95}$(Ti,Cr,Nb,Mo,W)$O_3$ structural and chemical characterization.** (a) X-ray diffraction patterns of films grown on different substrates in the 35-40 nm thickness range, (b) X-ray diffraction patterns of films grown on LSAT substrate with different thicknesses from 20 nm to 54 nm, (c) the 35 nm film on LSAT reciprocal space map of the (-103) reflections, (d) 35 nm film on LSAT SAED pattern with insets highlighting the film and substrate diffraction spots (acquired from the [100] zone axis), (e) drift-corrected HAADF image of the same sample showing uniform cubic perovskite structure with a zoomed in inset, see Figure S11 for annular bright field STEM image, and (f) representative EDX map showing the random distribution of B-site cations. The EDX maps for the individual elements and their corresponding STEM image are included in Figure S12. Note that (110) $DyScO_3$ and (110) $GdScO_3$ are equivalent to (001) pseudocubic orientation.

Figure 2(d) shows selected area diffraction (SAED) for the 35 nm film, this pattern is most consistent with perovskite structure with a longer, on average, out-of-plane lattice parameter than the substrate (Figure S11). Figure 2(e) shows a cross-sectional high-angle annular dark-field (HAADF) image along the ⟨100⟩ zone. Intensity fluctuations indicate more disorder near the film interface but they do not indicate a regular pattern of misfit dislocations despite the 2.5% lattice mismatch. Figure 2(f) shows a STEM-EDX map highlighting *B*-site cations with no indications of chemical segregation on the few-nm length scale. Individual maps for each cation are shown in Figure S12 that further support this interpretation. Power-dependent second harmonic generation (SHG) measurements were collected as a final probe of structure and symmetry. The analysis (shown in Supplementary Note 7) does not show SHG activity thus indicating a long range centrosymmetry.

Many *B*-site cations present favor multiple valence states, especially when processed at high temperatures and oxygen-lean conditions. As such, the chemical disorder is likely accompanied by a highly charge disordered environment where neighboring cations adopt oxidation states other than $4^+$. A comprehensive set of X-ray photoelectron spectroscopy (XPS) scans are conducted to resolve and quantify this likely valence milieu. The most informative scans and their fitting results are summarized in Figs. 3(a-c), with additional supporting information available in Table S1 and Figure S15. The most important findings include: (i) Nb is approximately 82% $Nb^{4+}$, 11% $Nb^{5+}$, and possibly 7% $Nb^{2+}$ possibly suggesting *A*-site occupancy. It is important to note, however, that $Nb^{3+}$ cannot be excluded, as detailed in the Supplementary Notes 10 and 11 and Figure S18; (ii)

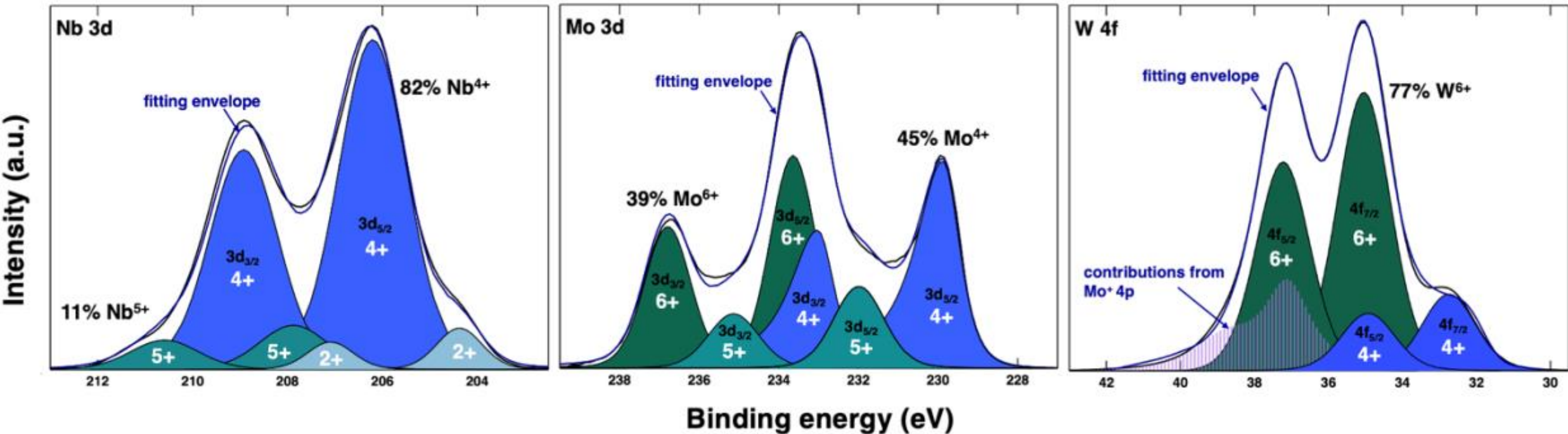

**Figure 3.** $Sr_{0.95}(Ti,Cr,Nb,Mo,W)O_3$ XPS fittings for (a) Nb 3*d*, (b) Mo 3*d* and (c) W 4*f* illustrating that the films exhibit magnified charge disproportionation. Notably, we cannot properly fit the W4*f* XPS spectra without accounting for Mo 4p peaks. This highlights the matrix effects present in such a chemically disordered system. Table S1 summarizes the oxidation states for all B-cations with the A-site being predominantly occupied with $Sr^{2+}$ (supplementary Notes 8 and 11).

Mo is 45% $4^+$, 39% $6^+$ and 16% $5^+$; (iii) W is 77% $6^+$ and 23% $4^+$; and (iv) Cr and Ti are $3^+$ and $4^+$ respectively. It is critical to acknowledge the valence quantification challenge in these complex mixtures. As shown in the Supplementary Notes 8-10, reference sample measurements and extensive fitting procedures were used to meet this challenge. For corroboration, X-ray absorption near edge structure (XANES) measurements were collected on a 200 nm film as detailed in Supplementary Note 11. XANES provides an effective average charge state between $5^{\pm}$ and $6^+$ for W, which is accessible due to its high absorption, consistent with our XPS results.

The entire valence determination experiment can be tested for self-consistency by applying charge neutrality. The details of this calculation are shown in Supplementary Notes 8, but the net result, when considering all measured valence values and the deliberate Sr vacancy concentration, is an equal and opposite positive and negative charge summation.

# 3 Electron transport properties under chemical disorder

As shown above, multiple cations must access multiple valence states in these $Sr_{0.95}BO_3$ crystals to maintain charge neutrality. This creates an 'electron scattering wilderness' superimposed on a lattice with remarkable crystalline fidelity, a combination that can promote flat temperature and frequency dependences to charge transport and optical responses respectively.

### *3.1 Optical Properties*

In Figure 4, we show $Sr_{0.95}BO_3$ optical properties and compare them with our PLD-grown $Sr_xNbO_3$ (Supplementary Note 9), which is reported to have the widest transparency window among end-members.[9,11,12] Figure 4(a) shows the real part of the dielectric function while Figure S19 (b-c) show the imaginary part of the dielectric function and the extinction coefficient, respectively, all calculated from ellipsometry, these values are generally consistent across all thicknesses, with modest variations likely arising from stoichiometry, strain relaxation, surface scattering, and multiple internal reflections. The real part of the dielectric function ($\varepsilon_1$) crosses zero at around 1.33 eV, this energy corresponds to a screened plasma frequency where ionic screening and interband electronic transitions introduce non-Drude contributions to the energy dependence.

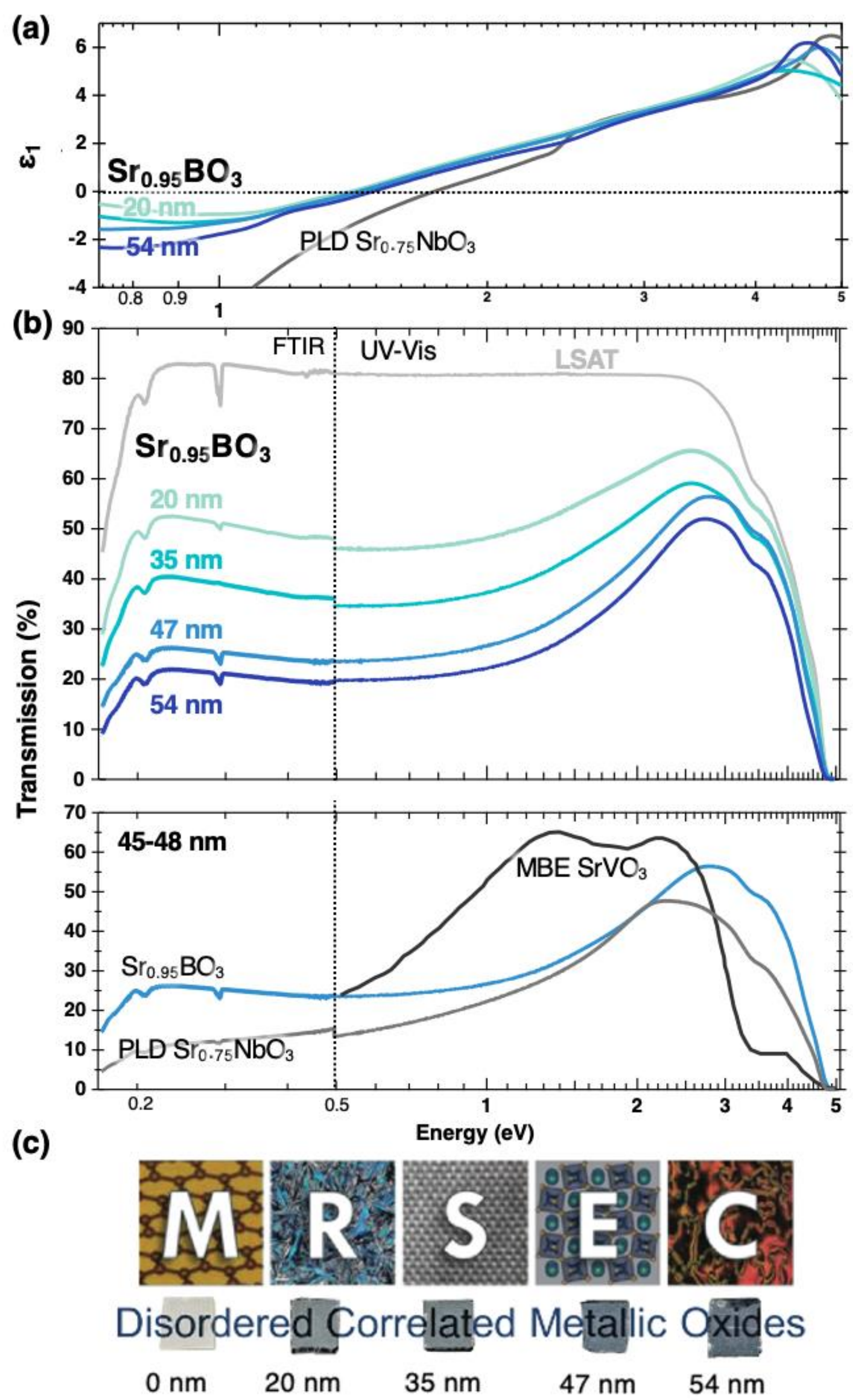

**Figure 4. $Sr_{0.95}BO_3$ optical properties.** (a) shows the real part of the complex dielectric function of $Sr_{0.95}BO_3$ in comparison to our 48 nm PLD grown $Sr_xNbO_3$; (b) upper part shows the corresponding optical transmission (%) from UV-Vis and FTIR, while the lower panel shows the transmission of 47 nm $Sr_{0.95}BO_3$ film in comparison to our $Sr_{0.75}NbO_3$ and Zhang et al. MBE $SrVO_3$ with comparable thicknesses; and (c) depicts the real samples photograph of $Sr_{0.95}BO_3$ film series.

This is comparable to the situation for $SrVO_3$, and red-shifted compared to $Sr_xNbO_3$ and $SrMoO_3$ [2,9,12]. In the context of perovskite conductors, the imaginary part ($\varepsilon_2$) and the extinction coefficient (k) of $Sr_{0.95}BO_3$ are quite low in the visible and UV regimes, with a sharp increase at higher photon energies attributed to $E_{O2p\text{-}Bdt2g}$ interband transitions. Figure S19(d) shows k calculated from ultraviolet-visible (UV-Vis) spectroscopy, closely matching ellipsometry results and reinforcing our confidence in the measurements and fittings.

Figure 4(b) shows transmission spectra for $SrBO_3$ as a function of film thickness from UV-Vis and Fourier-transform infrared spectroscopy (FTIR). $Sr_{0.95}BO_3$ films exhibit remarkable transparency in the IR and UV compared to the seminal molecular beam epitaxy (MBE) $SrVO_3$[2] and our PLD grown $Sr_xNbO_3$ with comparable thickness. The transparency enhancement in the UV is likely due to combined effects from Nb, Mo and W with large $E_{O2p\text{-}Bdt2g}$. We hypothesize that the IR transparency and the unusually flat IR dispersion observed in transmission and the real part of the dielectric function for $Sr_{0.95}BO_3$ arise from: (1) the increased effective mass resulting from correlation effects and electron-phonon coupling[26], and (2) the extreme chemical disorder, which suppresses intraband transitions by increasing electron scattering thus reducing low-energy transition probabilities. To further investigate this, we now examine the electrical properties.

### *3.2 Electrical Properties*

Figure 5 shows temperature-dependent resistivity of $Sr_{0.95}BO_3$, our PLD-grown $Sr_{0.75}NbO_3$, the seminal $SrVO_3$ MBE films of Zhang et al[2], and the highest conductivity $SrMoO_3$ single crystal from Nagai et. al[27]. The $Sr_{0.95}BO_3$ and $SrNbO_3$ film thicknesses are chosen for best comparisons to literature and both show a weak temperature dependence. In context, all $Sr_{0.95}BO_3$ end-members, except band-insulating $SrTiO_3$, are reported or predicted to be weakly metallic.[11,12] We attribute the weak $Sr_{0.95}BO_3$ temperature dependence to the combination of high crystalline fidelity, enhanced electron correlations, and simultaneous chemical disorder, which creates local metallic regions separated by a distribution of small transport barriers that reflect a nonperiodic or pseudoperiodic potential modulated by chemical disorder. Similar behavior is observed at all film thicknesses and on different substrates (Supplementary Note 15). The 20 nm $Sr_{0.95}BO_3$ film in Figure 5(a) reproducibly exhibits a room-temperature resistivity of ~ 460 μΩcm. However, small deviations from the optimized synthesis conditions can increase resistivity by more than 3×, with rapid quenching from the deposition temperature being especially critical. Supplementary Note 13 details establishing the deposition conditions for achieving high-conductivity films.

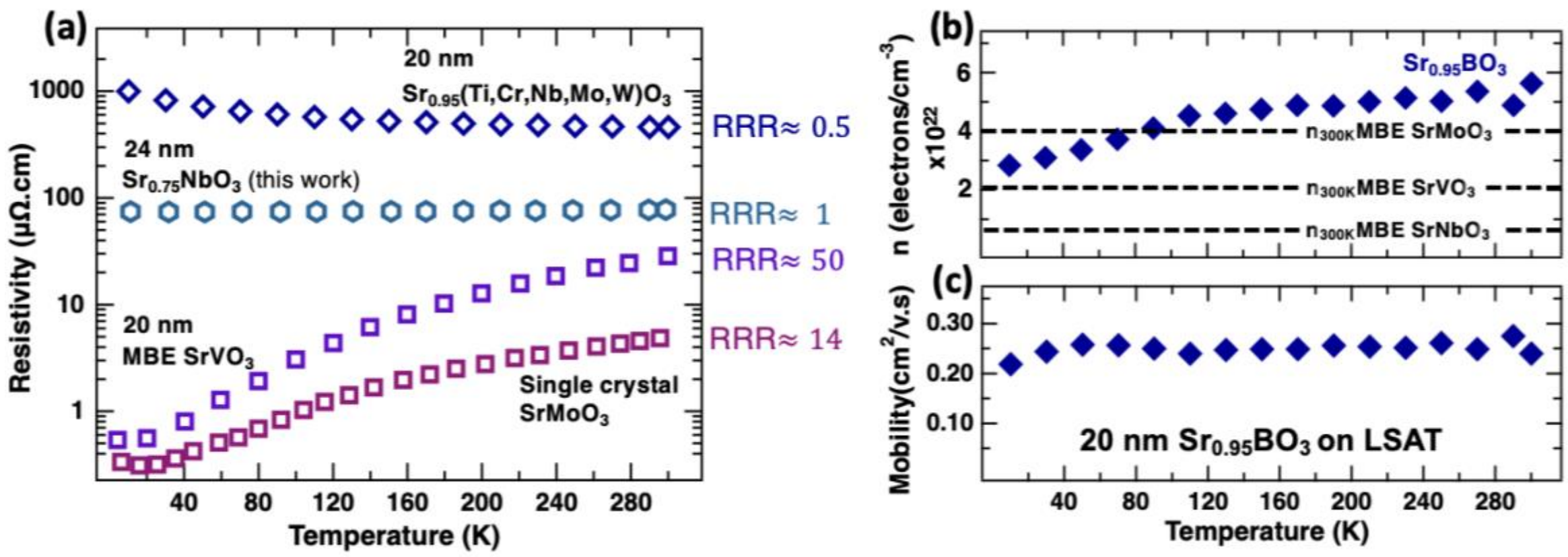

**Figure 5. $Sr_{0.95}BO_3$ electrical properties.** (a) Temperature-dependent resistivity for $Sr_{0.95}BO_3$ compared to that of MBE $SrVO_3$, single-crystal $SrMoO_3$, and our PLD $Sr_{0.75}NbO_3$ (b) carrier concentration and (c) mobility in a 20 nm $Sr_{0.95}BO_3$ film on LSAT as a function of temperature.

Both mobility and carrier concentration of $Sr_{0.95}BO_3$ films exhibit a comparatively flat temperature dependence, both contribute directly to the negative temperature-dependent resistivity coefficient and the small residual resistivity ratio (RRR) compared to end-members. The RRR decrease from $SrVO_3$ to $Sr_{0.75}NbO_3$ to $Sr_{0.95}BO_3$ in Figure 5(a) follows the trend of increasing chemical disorder while Figure 5(b) shows that $Sr_{0.95}BO_3$ has a room temperature carrier concentration ~ 5 $\times 10^{22}$ $cm^{-3}$ which is 2X and 4X larger than $SrMoO_3$[13] and $SrNbO_3$[9], respectively. We attribute this large value to high band filling from a large overall average *B*-site valence while maintaining a relatively small unit cell volume - unit cell volume is 1.6% and 4.6% smaller than $SrMoO_3$[13] and $SrNbO_3$[9], respectively. Conversely, the 0.22 $cm^2V^{-1}s^{-1}$ room temperature electron mobility in Figure 5(c) is significantly lower than that of end-members like $SrNbO_3$ (~8 $cm^2V^{-1}s^{-1}$)[9]. This reduced mobility, like our optical analysis, could be attributed to the increased scattering caused by disorder.

# 4 The evolution of short-range ordering

Periodic elemental trends and DFT can guide HEO formulation selection, but they assume chemical homogeneity (Section 1 and Supplementary Note 2). It is important to recognize that short-range chemical ordering, which is often overlooked, is very likely, particularly for these far-from-equilibrium compositions. To do so, we computationally examine the Warren-Cowley parameter ($w_{\alpha\beta,n}$) for the 1st and 2nd nearest neighbor shells in $Sr_{0.95}BO_3$ equilibrated at 1500 K with Metropolis Monte Carlo sampling, as illustrated in Figure 6 (See Supplementary Note 14 for formal definitions and comment on temperature).[28,29] For individual $\alpha$-$\beta$ pairs in the $n^{th}$ shell, negative $w_{\alpha\beta,n}$ values indicate clustering tendencies while positive values indicate repulsion tendencies. The most interesting result is that Ti-containing pairs have the smallest $w_{Ti\beta,n}$ values, suggesting that Ti atoms are adaptable to many local environments and serve as a "mixing" agent. This supports our Section 1 hypothesis that Ti enhances the stability in the cubic perovskite structure. Additionally, and most generally, cation pairs that are either highly attractive or repulsive in the first shell have the opposite interaction in the second shell, with similar relative magnitude. This suggests an overarching tendency to avoid cation clustering at the polyhedral level, but a possibility for short range ordering at the several-polyhedral length scale. Both are consistent with the present models that explain the electrical and optical properties. We note that Sr vacancies and multiple cation valence states are not included in the present simulations, future studies that self-consistently incorporate both during Monte Carlo runs will represent these atomic-scale interactions with higher accuracy.

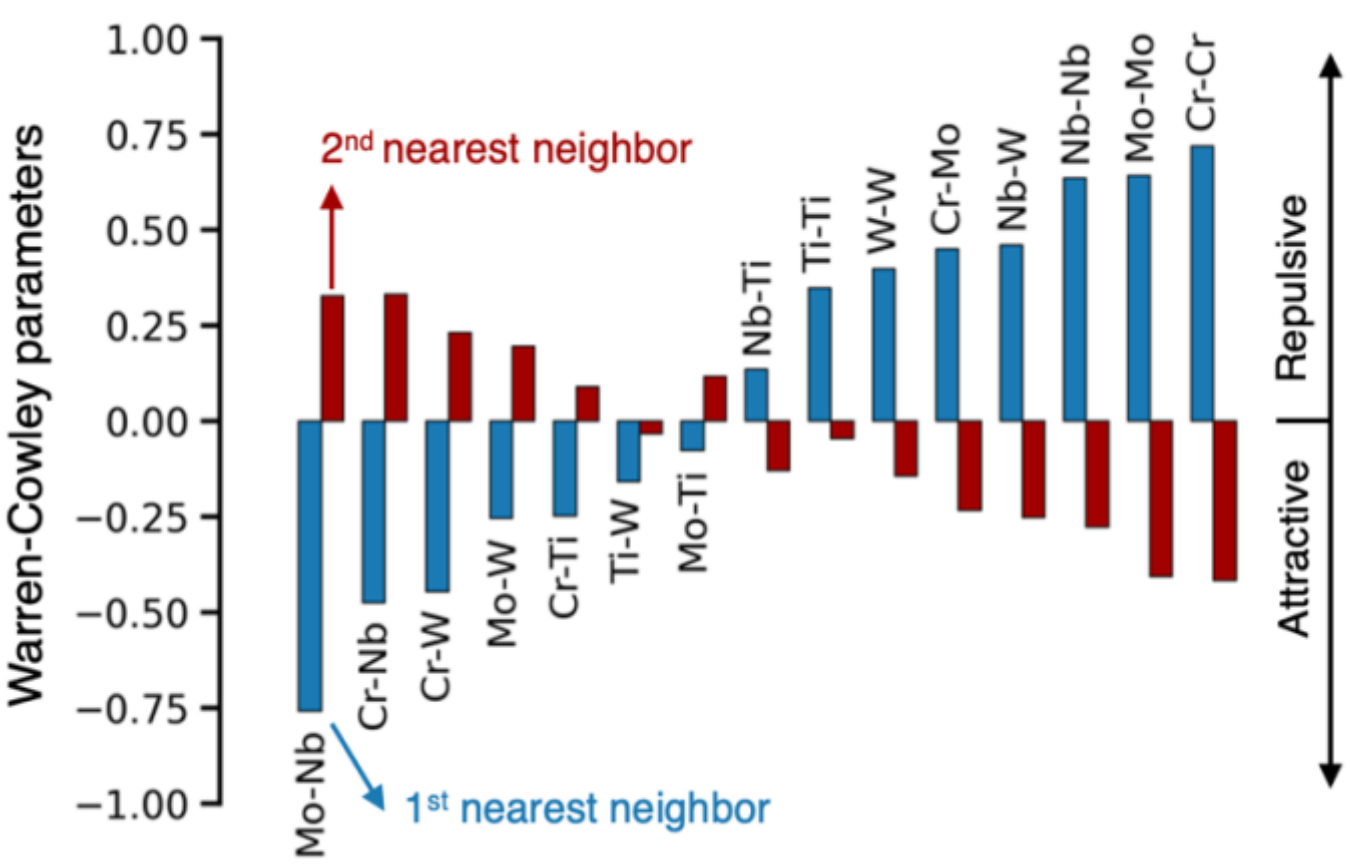

**Figure 6. $Sr_{0.95}BO_3$ short range ordereing.** Warren-Cowley (WC) parameters for pairs of B-cations in the first and second nearest neighbor shells equilibrated at 1500K. Positive WC parameters indicate repulsion; negative WC parameters indicate attraction.

## Outro

Colloquially, “disorder” connotes chaos, but within crystalline materials it can often be a source of functionality, as in optical glasses, relaxor ferroelectrics, superconductors, and hybrid lead–halide perovskites.[30–35] HEOs provide a new landscape to engineer functionality through chemical disorder by preserving long-range translational order (enabling analyses based on Bloch’s theorem) and local connectivity (ensuring consistent coordination geometries).[18,20,22,23] Extending the high-entropy design concept to correlated oxides, we engineer $Sr_x(Ti,Cr,Nb,Mo,W)O_3$ as a pioneering chemically disordered and positionally ordered correlated metallic oxide. Electron correlation enriches the complexity of disordered materials, intertwining electrical, optical, and magnetic responses, which make disordered correlated oxides appealing for nonlinear transport and quantum applications.[7,25,36] Targeted functionality composition design – such as adding Fe on the B-site for ferrimagnetism[37] or Cu for superconductivity[38] – unlocks a new composition frontier awaiting discovery, exploration, and implementation.

## Acknowledgments

The authors acknowledge the support from the Penn State Materials Research Science and Engineering Center *Center for Nanoscale Science* under National Science Foundation award DMR-2011839. DFT calculations utilized resources from the Roar Collab cluster of the Penn State Institute for Computational and Data Sciences. The authors also acknowledge the help from the XPS lab and Jeff Shallenberger at Penn State MCL. DFT calculations used to parametrize the cluster expansion model were carried out on the Rockfish cluster at the Advanced Research Computing at Hopkins (ARCH) core facility, through allocation PHY220113 from the Advanced Cyberinfrastructure Coordination Ecosystem: Services & Support (ACCESS) program, which is supported by National Science Foundation grants #2138259, #2138286, #2138307, #2137603, and #2138296. The authors would also like to acknowledge many helpful and enlightening discussions with Zhiqiang Mao, Long-Qing Chen, Susan Trolier-McKinstry, George N. Kotsonis, Bed Poudel, Rui Zu, Jackson Spurling, Sarah Boehm and Francisco Viera regarding different nuances of the present study.

# Methods and Supplementary Information

## Experimental Methods

Sample Synthesis $Sr_xBO_3$ were deposited on different substrates by PLD using a Coherent 248 nm KrF laser ablating sintered ceramic targets (supplementary note 5). Substrates were cleaned with methanol followed by a 1hr- UV-ozone treatment then were adhered to a heater buck with silver paint (Ted Pella Leitsilber 200). The substrates were transferred to the heater at 850°C and growth started 3 minutes after transfer. 40 sccm Argon was flowed into the chamber, the gate valve was throttled until the chamber pressure stabilized at 50mTorr. Laser fluence was 1.4 J/cm$^2$ with a total laser pulse energy of 110 mJ. All films were quenched to ambient lab conditions in less than 6 minutes after the final laser pulse to mitigate post-synthesis annealing effects (See supplementary Note 13 for more details).

X-ray diffraction (XRD) was performed using a PANalytical Empyrean diffractometer with Bragg-Brentano HD (BBHD) incident beam optics. We employed a 2-bounce Ge hybrid monochromator for the high-resolution scans and reciprocal space maps (RSMs). The primary diffracted beam optic was a programmable divergent slit and a PIXcel$^{3D}$ detector combination.

X-ray photoelectron spectroscopy (XPS) experiments were performed using a Physical Electronics VersaProbe III instrument equipped with a monochromatic Al kα x-ray source (hν = 1,486.6 eV) and a concentric hemispherical analyzer. Charge neutralization was performed using both low energy electrons (<5 eV) and argon ions. The binding energy axis was calibrated using sputter cleaned Cu (Cu $2p_{3/2}$ = 932.62 eV, Cu $3p_{3/2}$ = 75.1 eV) and Au foils (Au $4f_{7/2}$ = 83.96 eV). Peaks were charge referenced to $CH_x$ band in the carbon 1s spectra at 284.8 eV and in some cases to Sr $3d_{5/2}$ at 132.9 eV. Measurements were made at a takeoff angle of 85° with respect to the sample surface plane. This resulted in a sampling depth of roughly 4.5-9.5 nm (95% of the signal originated from this depth or shallower). Quantification was done using instrumental relative sensitivity factors (RSFs) that account for the x-ray cross section and inelastic mean free path of the electrons. On homogeneous samples major elements (>5 atom%) tend to have standard deviations of < 3% while minor elements can be significantly higher. The analysis size was ~200µm in diameter. All fittings were performed in CasaXPS.

Transmission Electron Microscope Sample preparation for S/TEM studies were carried out using Scios 2 Focused Ion Beam (FIB). The cross-sectional specimen was extracted at 30kV ion beam and thinned at 5kV ion beam. Finally, 2kV ion beam was used to clean the sample. Selected Area Electron Diffraction experiments were performed on Talos X2 at 200 kV accelerating voltage. The selected area aperture was placed over both the film and substrate. STEM and STEM-EDX studies were carried out at a 300 kV accelerating voltage on the aberration-corrected Titan G2 microscope at MCL, Penn State. Drift correction on HAADF-STEM image was performed using open-source MATLAB code on images acquired in orthogonal scan directions.[1]

Ellipsometry spectra in (Ψ,Δ) were collected using three different ellipsometry. J.A.Woollam M-2000 Ellipsometer was used at the incident angle of 45° to 85° for the spectral range of 0.734 to 5.042 eV. J.A.Woollam M-2000F Focused Beam Ellipsometer was used at the incident angle of 64.54° for the spectral range of 1.240 to 6.458 eV. Prior to the measurement of SrBO3 samples, the ellipsometry spectra of LSAT substrate were measured for the modeling. The data were modeled by B-spline fitting with Kramers-Kronig relations using J.A. Woollam CompleteEASE software.

Ultraviolet-Visible-near-IR Spectroscopy The transmission (%) and reflection (%) measurements in the wavelength range from 250 to 2500 nm (4.959 to 0.4959 eV) were measured using an Agilent Cary5000 Ultraviolet-Visible-near-IR Spectroscopy with an integral sphere. The transmission intensity of a sample was normalized by the intensity without a sample in the same optical path. The reflection intensity of a sample was normalized by the intensity of the Spectralon, the reflectance standard.

Fourier-Transform Infrared (FT-IR) Spectroscopy The transmission (%) and reflection (%) measurements in the wavelength range from 2500 to 16000 nm (0.4959 to 0.07749 eV) were performed using a Bruker Hyperion 3000 Microscope with a 15× objective lens. The transmission intensity of a sample was normalized by the intensity without a sample in the same optical path. The reflection intensity of a sample was normalized by the intensity of the gold film on the Si substrate.

Temperature dependent Hall and resistivity measurements were performed in a Physical Properties Measurement System (PPMS). The samples were configured in the standard Van der Pauw geometry. A baffle rod fixed with a Au-coated sealing disc hovered <1 cm above the sample to ensure thermal equilibration; additionally, the sample was held at each temperature for the same purpose. The thermal rate between measurements was set to 5 K/min. Sheet resistance and Hall effect measurements were performed using a Keithley 2450 Source Measure Unit and a Keithley 3706A-S to switch amongst all to Van der Pauw configurations. The source current ranged from 100 to 200 μA. The magnetic field was swept between ±2 T for Hall voltage measurements on LSAT. We do not report the Hall measurements as a function of temperature for the films on GSO and DSO due to their strong paramagnetic background.

## Supplementary Notes

Note 1: Clarifying Terminology and Thermodynamic Principles

Chemical disorder, high entropy and entropy stabilized

In this study, we utilize the term "High-Entropy Oxide" (HEO) to describe crystalline oxide solid solutions characterized by four- or five-component near-equimolar cation mixtures occupying one or more equivalent lattice sites. This yields a large configurational mixing entropy, $S_{config}$. From a crystallographic viewpoint, a loosely defined crystal lattice still exists on average, with an indeterminate identity atom occupying each lattice site. We use the term "end-members" in reference to the constituent extremes of the compositional range in HEO solid solutions – each endmember has one determinate atom occupying equivalent lattice sites.

The terms "entropy" and "disorder" are often used interchangeably in the literature, though there is a subtle difference between them. "Disorder" implies a focus on a single configuration, whereas "entropy" encompasses a class of configurations. When describing HEOs as disordered, we refer to the inherent local asymmetries within the crystal in a "trapped" state which give rise to their unique properties and functionality. Simultaneously, the significant entropy gain from chemical mixing facilitates HEO phase formation, as the large $S_{config}$ dominates the entropy of mixing ($\Delta$s) and boosts the chemical potential change ($\Delta\mu = \Delta h - T\Delta s$) associated with forming a multi-component solution from the end-members, $h$ is molar enthalpy and $T$ is temperature. "Entropy-stabilized" is a more specific term introduced by Rost et. al that implies that $\Delta h > 0$ and that a critical thermodynamically determined temperature exists, above which entropy drives phase formation by making the overall $\Delta\mu$ negative.

Pulsed Laser Deposition (PLD) and Kinetic Stabilization

Pulsed laser deposition (PLD) often employs non-equilibrium kinetics and involves condensing precursors from a high-entropy initial state, allowing kinetic stabilization of a broader spectrum of atomic and electronic configurations whose bulk synthesis may require extreme physical or chemical conditions. The high-entropy phase is subsequently trapped at room temperature with unique local asymmetries and fluctuations in chemical, structural, and electromagnetic order parameters. We have shown in previous studies that controlling PLD growth conditions allows for controlling cations' oxidation states and for subtle nanoscale microstructural reconfiguration.

Correlation and disorder

To better appreciate correlation's role under disorder, we examine two extreme hypothetical cases from the electrons' perspective: one with metallic components exhibiting little to no electron correlation, and the other with insulating oxides where electrons are fully localized in closed electronic shells. In the former case, free electrons will screen chemical disorder, minimizing its impact on transport properties apart from a likely reduction in electronic conductivity. In the latter case, disorder may primarily influence the optical behavior. By permitting oxygen vacancies and cation multivalency, however, disorder can introduce unique magnetic phases, electrochemical activity, ionic conductivity, and defect-mediated hopping conduction in insulating oxides.[2–7]

The intermediary case of correlated electron end-members, as presented in the main manuscrpt, therefore offers an ideal setting for investigating and engineering complex electron behaviors and interactions. Disorder in these systems amplifies correlated electron interactions, intertwining electrical, optical, and magnetic responses. This fuels interest in disordered correlated oxides for developing various quantum applications and devices, even beyond conventional linear transport engineering including disorder induced nonlinear and high-order hall effects.[8–10]

## Note 2: DFT calculations

### Magnetic ordering and symmetry braking surpass naïve DFT in capturing correlation

While long-range symmetry provides a starting point for modeling electron-correlated HEOs, the enormous number of atoms necessary to accurately model these disordered systems as well as the lack of analytical solutions to fully account for correlation poses significant challenges to our theoretical and computational predictability. Nonetheless, straightforward DFT calculations coupled with periodic trends offer a wealth of crystal chemistry rules to inform our composition selection, as outlined in Section 1. Despite being common in literature, the so-called "naïve" DFT that utilizes only the cubic, nonmagnetic (NM) unit cell for perovskites omits and underestimates correlation compared to the more realistic, paramagnetic (PM) material, as has been recently shown for a variety of different perovskites[43]. Allowing colinear spins in an anti-ferromagnetic ordering (AFM-G) for 2x2x2 supercells in which all spins have maximum spin dissimilarity, however, results in an overestimate of correlation[11,12] . While neither method fully encapsulates the true PM behavior at room temperature, we consider both NM and AFM-G configurations in this study to provide a more comprehensive computational view for the explored $SrBO_3$ end members. We find that this provides valuable insights for designing and understanding our champion composition. The Vienna Ab-initio Software Package (VASP) 6.4.1 is used for DFT calculations with the projector augmented wave pseudopotentials v54.[13] The regularized-restored strongly constrained and appropriately normed ($r^2$SCAN) functional is used for its improved accuracy, hence here we did not consider the use of Hubbard U corrections.[14] Optical calculations are performed using the independent particle approximation for calculating the frequency-dependent dielectric matrix, in which the number of bands was doubled to ensure convergence of the energy spectrum. A Γ-centered k-point mesh of 8x8x8 was used for the 5-atom unit cells and scaled linearly with the size of the supercell. The k-point mesh was tripled (i.e. Γ-centered 24x24x24 for the unit cell) for band structure and optical calculations. To allow for symmetry breaking distortions (such as octahedral tilting), we rattle all atoms prior to starting the relaxation process; forces were minimized to less than 10 meV/Å, and the global cubic symmetry was maintained. Pymatgen[15], the Atomic Simulation Environment[16], and SUMO[17] were used for analysis. The unscreened plasma frequency was extracted from calculated frequency dependent dielectric matrices calculated using the LOPTICS = True flag in VASP while $E_{O2p\text{-}t2g}$ gaps were determined using the atom-projected density of states. For completeness, we include below in Figure S1 the heat maps for the NM calculations for Figures 1(c), 1(e), and 1(g). Figure S2 presents a complete set of NM band structures for the relevant end members. We outline the relevant band structure features using the NM unit cell calculations as this provides a clear demonstration of the bands features compared to AFM-G. However, supercell AFM-G calculations are utilized for dielectric quantities as these allow for structural and magnetic symmetry-breaking motifs that have shown to be important in accurately predicting these quantities as outlined below.

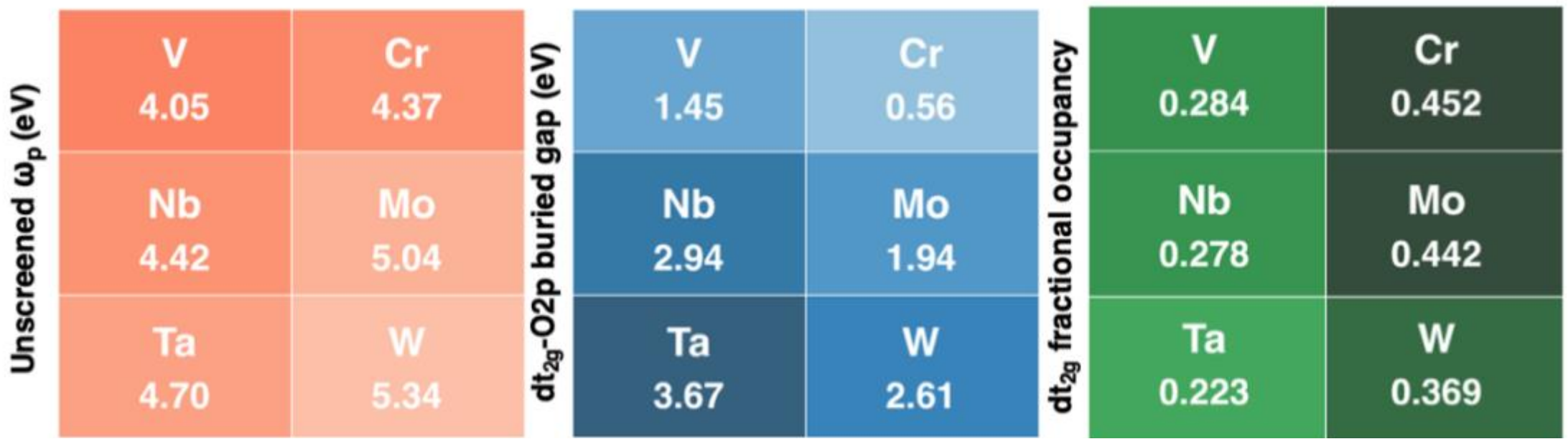

**Figure S1.** DFT nonmagnetic calculations heat maps for unscreened plasma frequency, $d_{2g}$-$O_{2p}$ buried gap and $dt_{2g}$ fractional occupancy.

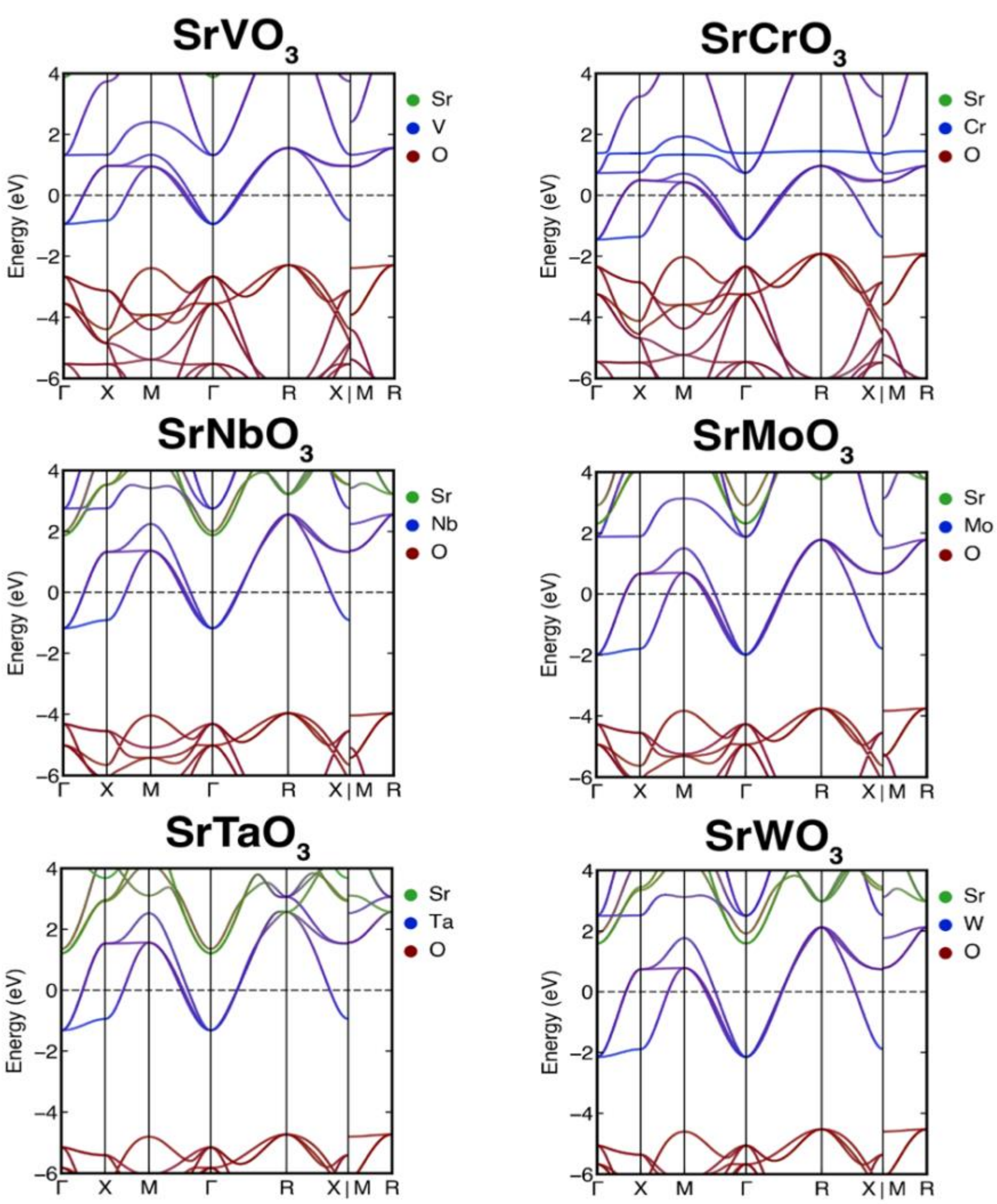

**Figure S2.** $SrBO_3$ NM end-member band structures

## DFT endmembers optical absorption

Figure S3 presents the optical absorption spectra of all relevant end-members, calculated from DFT using AFM-G ordering. $SrNbO_3$ and $SrTaO_3$ exhibit the lowest absorption in the UV and visible ranges, while $SrTiO_3$ shows the lowest absorption in the visible and infrared ranges. $SrVO_3$ and $SrCrO_3$ demonstrate better transparency in the visible range compared to $SrMoO_3$ and $SrWO_3$, and they outperform all other 5B and 6B elements in the lower energy ranges.

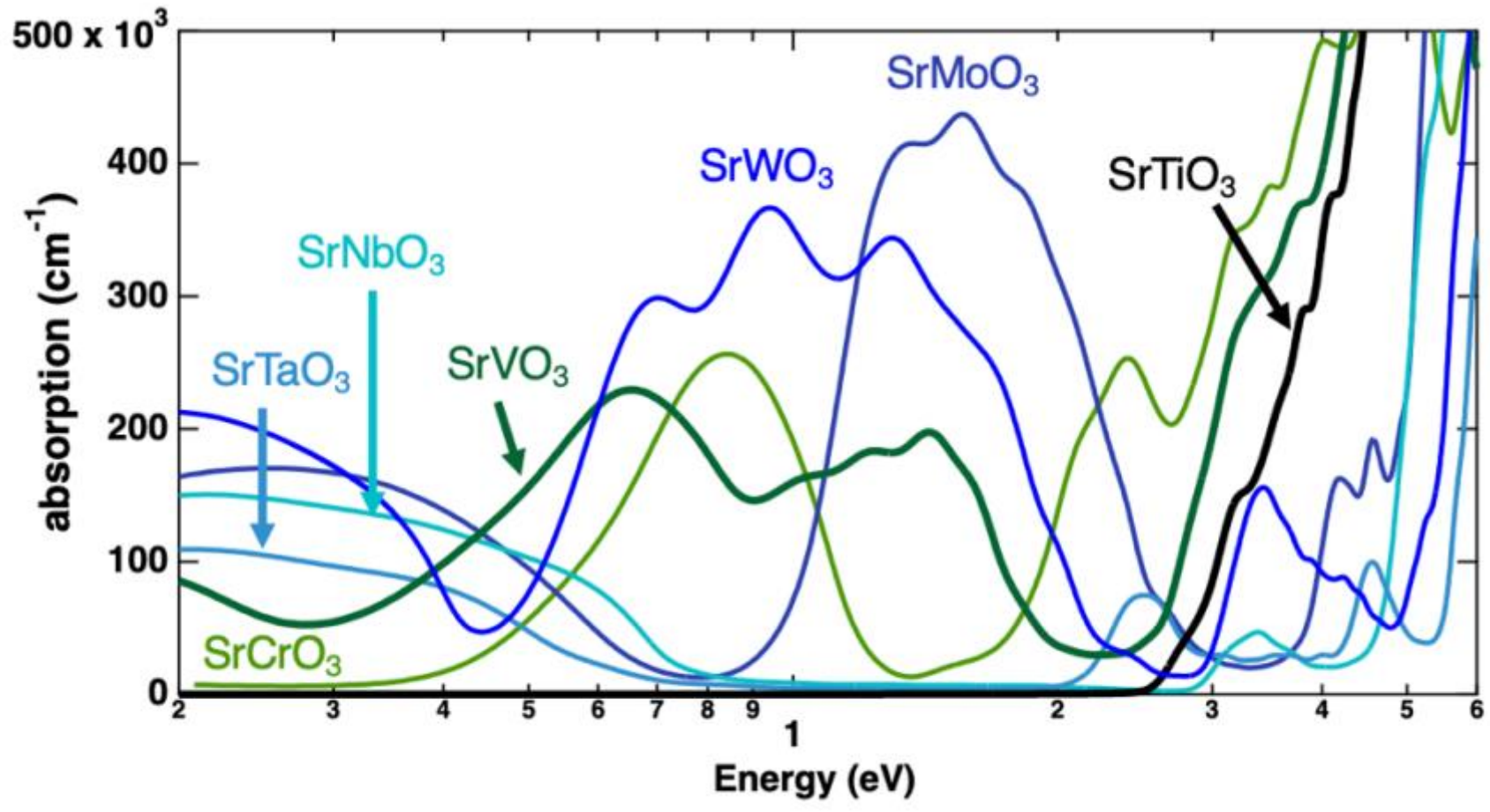

**Figure S3.** $SrBO_3$ AFM-G end-members optical absorption

## Note 3: Preparing the ceramic target

Bulk ceramics were prepared by mixing binary oxide powders to give stoichiometric solid solutions, namely SrO, $TiO_2$, $Cr_2O_3$, $Nb_2O_5$, $MoO_3$ and $WO_3$ purchased from Millipore-Sigma. Powders were checked for phase and stoichiometry at every processing step using X-ray diffraction (as discussed in Methods) and PANalytical Epsilon 1 X-ray fluorescence. We reacted the B-cations first at 650$^o$C for 24 hours to avoid $MoO_3$ sublimation and to ensure proper B-site cations mixing. Then we added SrO to the B-cations mix at ambient conditions and shaker-milled the mix with 5 mm and 3mm diameter yttrium-stabilized zirconia milling media for 18 hrs. Subsequently, we reacted the mix at 1200$^o$C for 24 hours to form $Sr_xBO_\delta$. Then we milled the reacted powder with 5 mm and 3mm diameter yttrium-stabilized zirconia milling media for another 18 hrs. Following that, we pressed the powder uniaxially into a 1" diameter pellet at 140 MPa (Carver Laboratory Press). The target pellet was then sintered in air at 1400°C for 18 hrs and air-quenched by direct extraction from the hot zone of the furnace. The XRD and SEM EDX of the bulk ceramic with 5% Sr vacancies are shown in Figure S4. The target exhibits two phases: one reminiscent of the perovskite structure and the other of the scheelite structure. From EDX maps, Ti, Cr and Nb seem to cluster in the same grains while Mo is well dispersed in the sample.

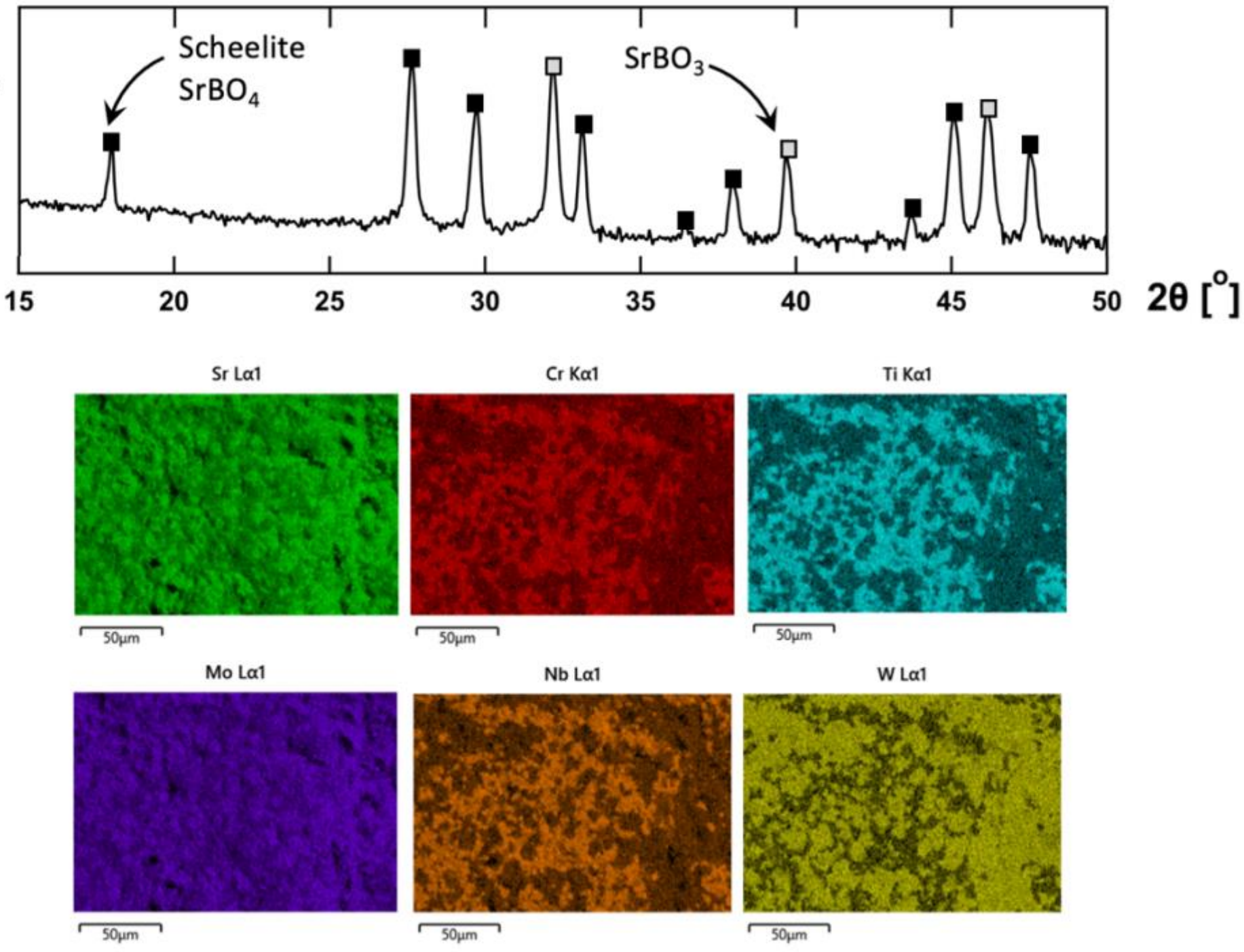

**Figure S4.** $Sr_{0.95}BO_3$ ceramic target powder diffraction and EDX maps

<u>Note 4: Vacancies on the A-site</u>

Inspired by previous work on $Sr_xNbO_3$[18], we investigated introducing Sr-vacancies on the A-site when preparing the PLD targets. Among those, films grown from $Sr_{0.95}BO_3$ consistently have the highest crystalline quality and lowest electrical resistivity. Future systematic experiments are needed to elucidate the role that vacancies play in the overall behavior of this perovskite family.

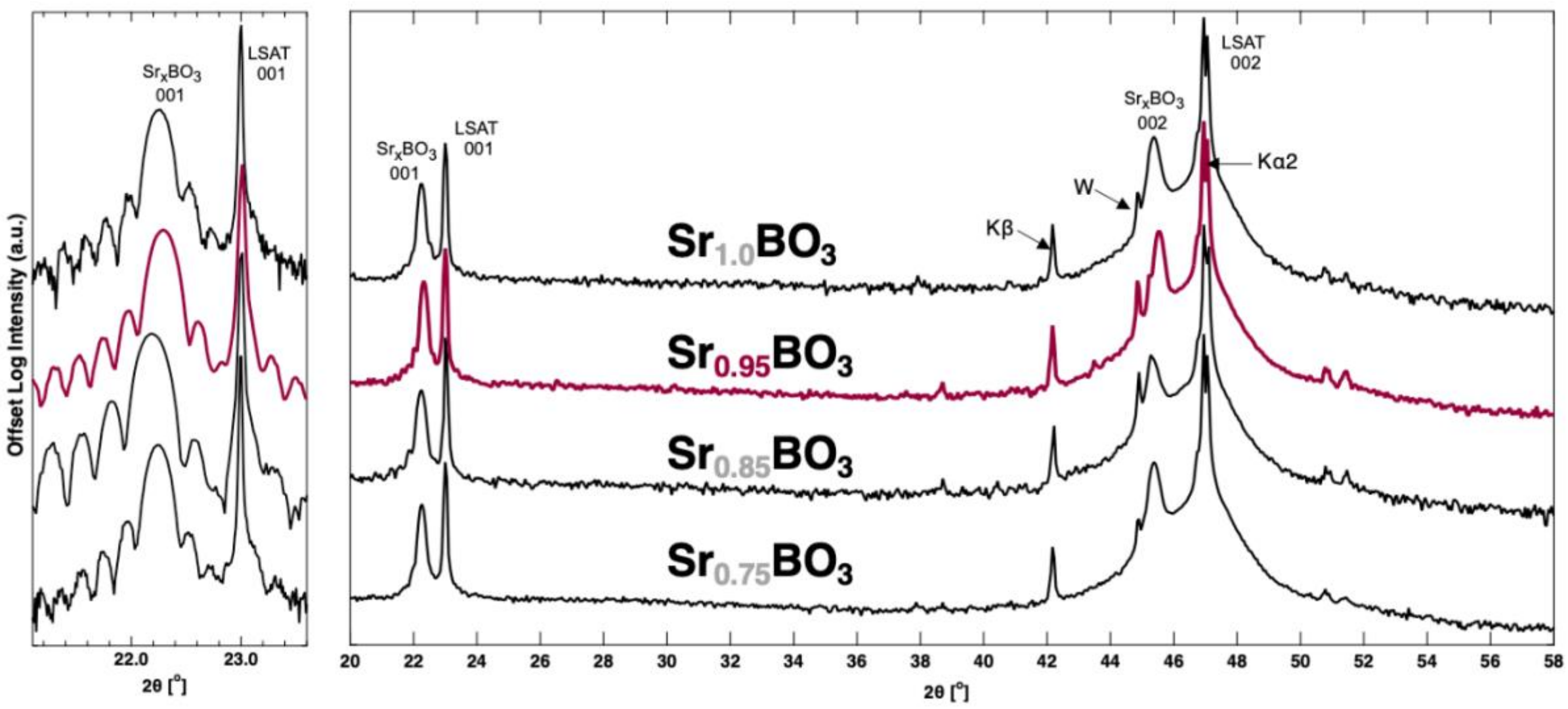

**Figure S5.** BBHD and high-resolution X-ray diffraction patters for films grown from a series of targets with varying A-site vacancies

Note 5: Growth on different commercially available substrates

Figure S6 depicts wide-angle 2θ−ω XRD scans for $Sr_{0.95}BO_3$ films grown on different technologically relevant substrates: (001) $(LaAlO_3)_{0.3}(Sr_2TaAlO_6)_{0.7}$ (LSAT), (001) $SrTiO_3$, (110) $DyScO_3$, (110) $GdScO_3$ and (001) $KTaO_3$, corresponding to Figure 2a. The surface morphology of the same set of films was characterized using an Asylum MFP3D atomic force microscope (AFM) in tapping mode as shown in Figure S7. AFM images indicate comparable smooth surfaces on all relevant substrates in this study. The temperature dependent resistivity for another set of films with 20 nm thickness are reported in Figure S8. The 20 nm films were chosen to facilitate comparison with existing literature (see Supplementary Note 13).

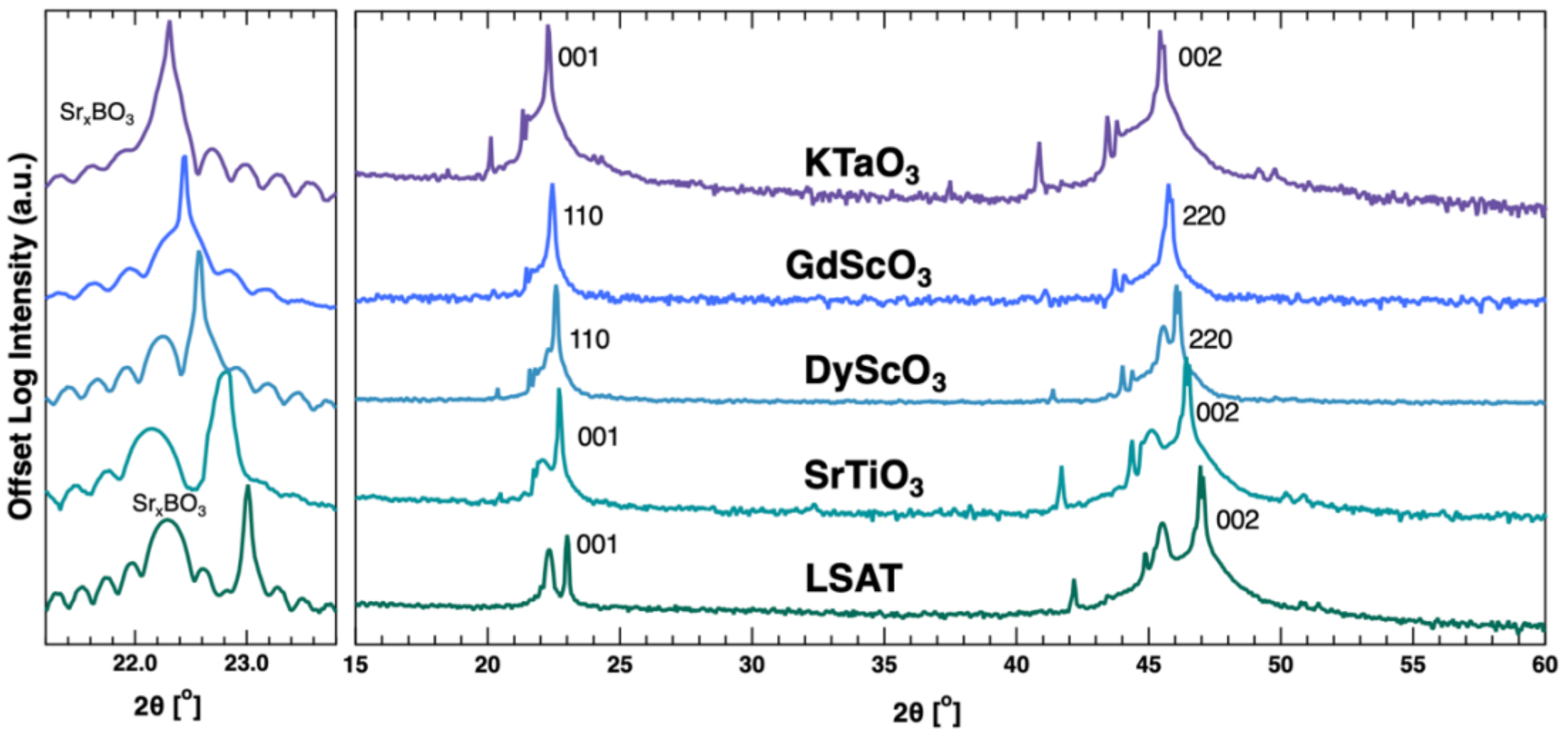

**Figure S6.** BBHD and high-resolution X-ray diffraction patters for $Sr_{0.95}BO_3$ 35 nm films grown on different substrates

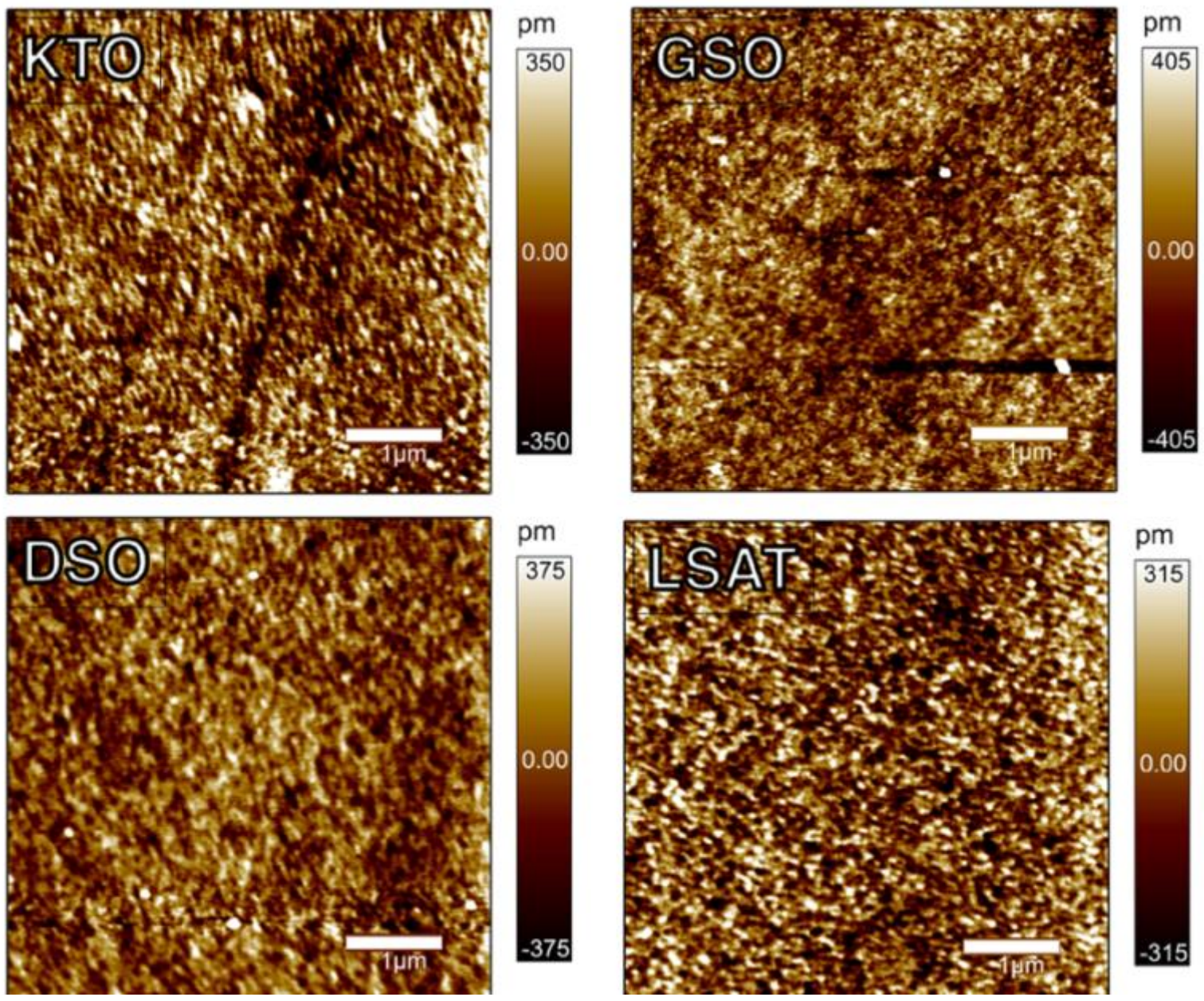

**Figure S7.** $Sr_{0.95}BO_3$ 35nm films AFM images

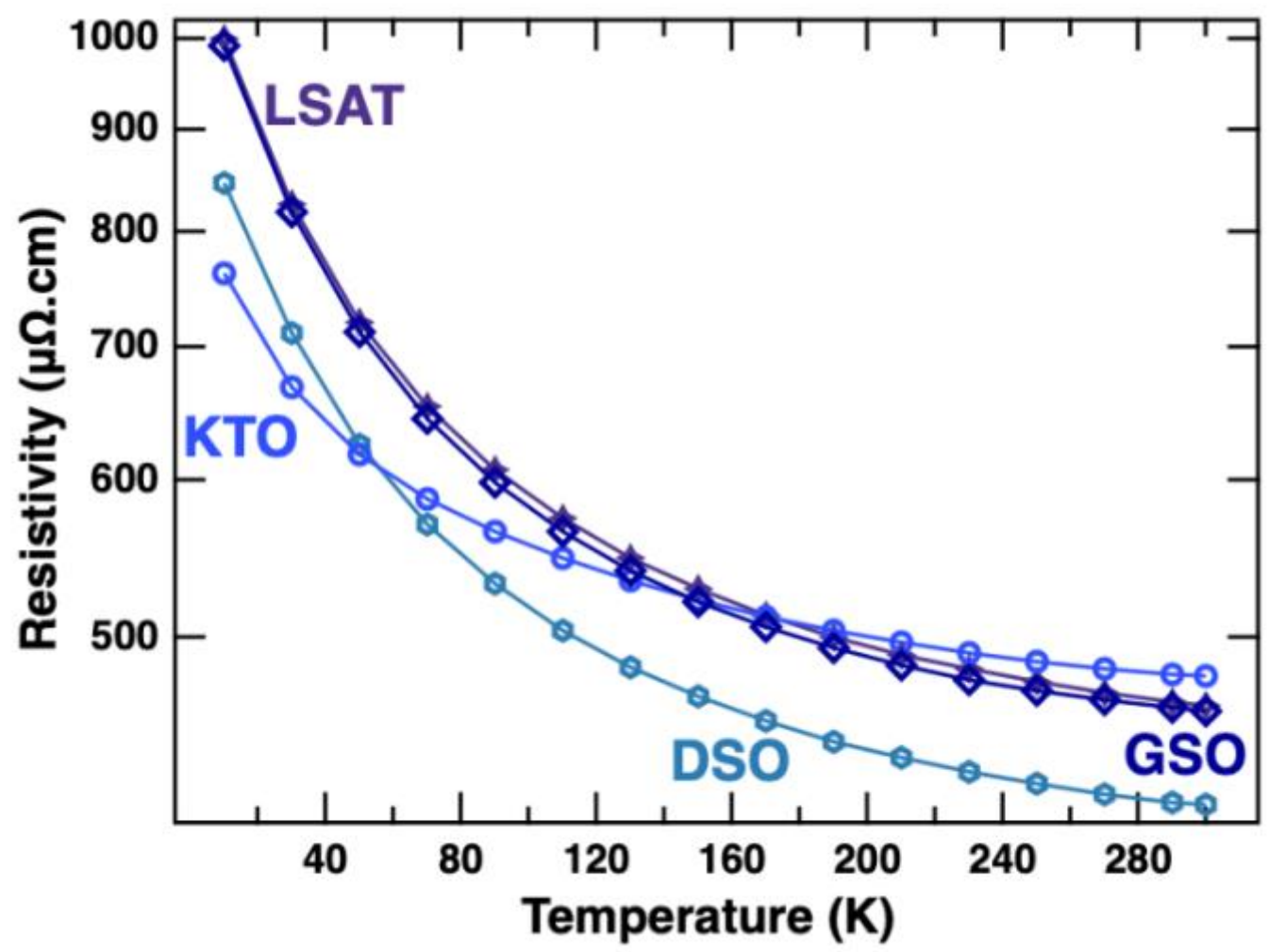

**Figure S8.** Temperature dependent resistivity of $Sr_{0.95}BO_3$ 20 nm films grown on different substrates

Note 6: Thickness Series on LSAT

Even though the films have the highest lattice mismatch with LSAT compared to other substrates, LSAT substrates are economic, are non-magnetic at low temperatures, have inherently large ultraviolet transmittance, remain insulating and do not react with the films. These characteristics make them suitable for measuring the films' electronic, optical, and magnetic properties with minimal disturbances. Therefore, we highlight $Sr_{0.95}BO_3$ thickness series grown on LSAT. Figure S9 presents wide $2\theta$- $\omega$ scans and the corresponding rocking curves ($\omega$ scans at constant $2\theta$) for the films presented in Figure 2b. Only peaks associated with 001 perovskite phase are present. The $\omega$ scans consist of a two-component superposition, one with a full-width-at-half-maxima (FWHM) of (0.18-0.37)° and the other with a FWHM of (0.028-0.045)°. This narrow spread in $\omega$ indicate very high-crystalline fidelity particularly in thicker samples. Furthermore, we sustain this high quality up to 300nm thickness as shown in Figure S10.

The TEM analysis in this study was performed on a 35 nm $Sr_{0.95}BO_3$ film on LSAT. Figure S11(a) represents the selected area corresponding to the electron diffraction in Figure 2d. The electron diffraction was acquired from both the HEO thin film and the LSAT substrate, and the lattice mismatch between LSAT and HEO is clearly observed in the diffraction pattern. Figure S11(b) presents an Annular Bright Field STEM image that highlights the positions of both oxygen atoms and cations, while Figure S11(c) shows the simultaneously acquired HAADF STEM image. Both STEM images illustrate a pristine cubic perovskite pattern. Bandpass filtering was applied to the atomic resolution STEM images in Figure S11 using Digital Micrograph. Figure S12 presents the EDX maps for each individual element, deconvoluting Figure 2f.

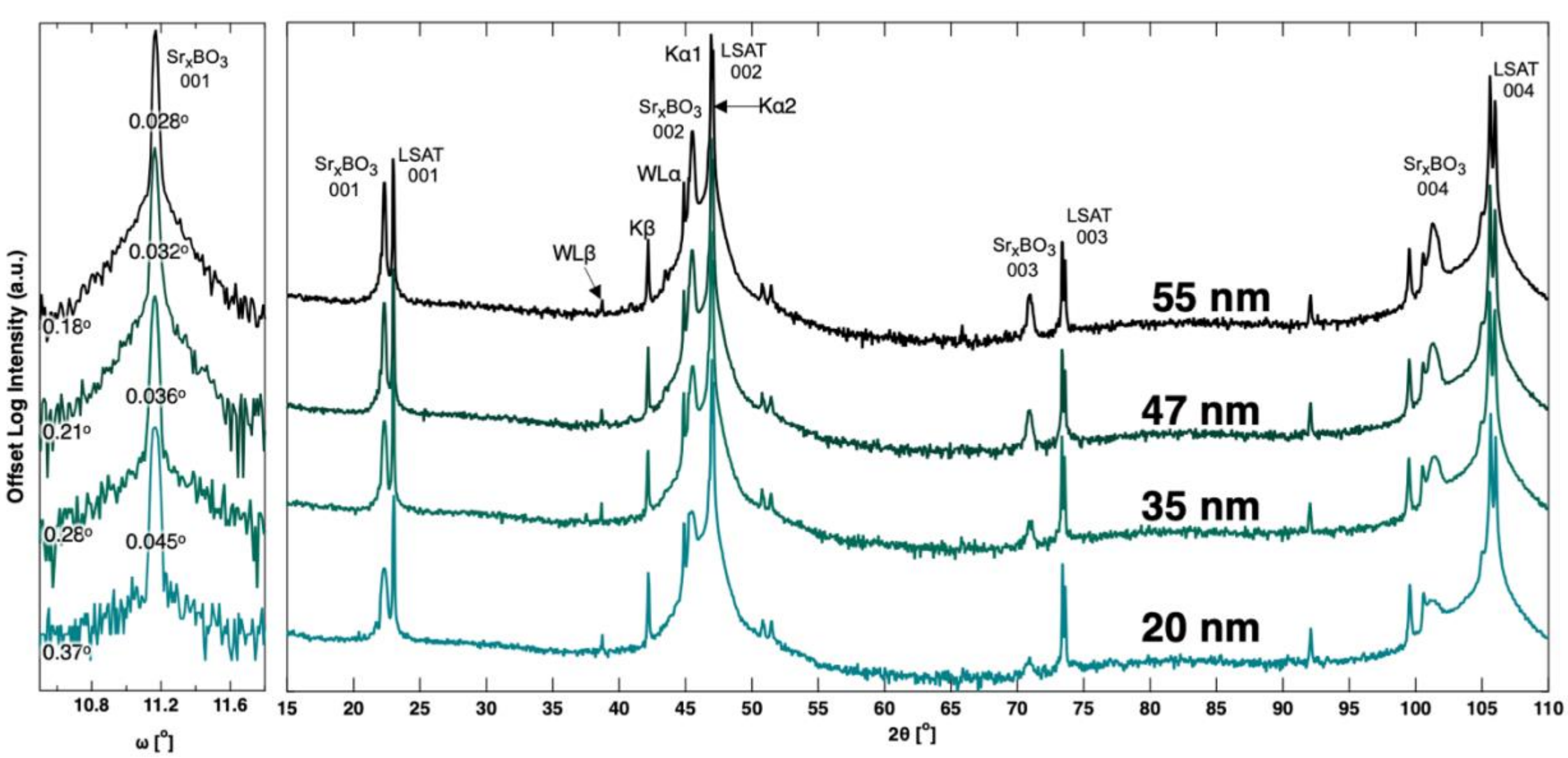

**Figure S9.** BBHD X-ray diffraction patterns and rocking omega curves for the $Sr_{0.95}BO_3$ films grown on LSAT from 20 nm to 55nm and shown in Figure 2 in the manuscript

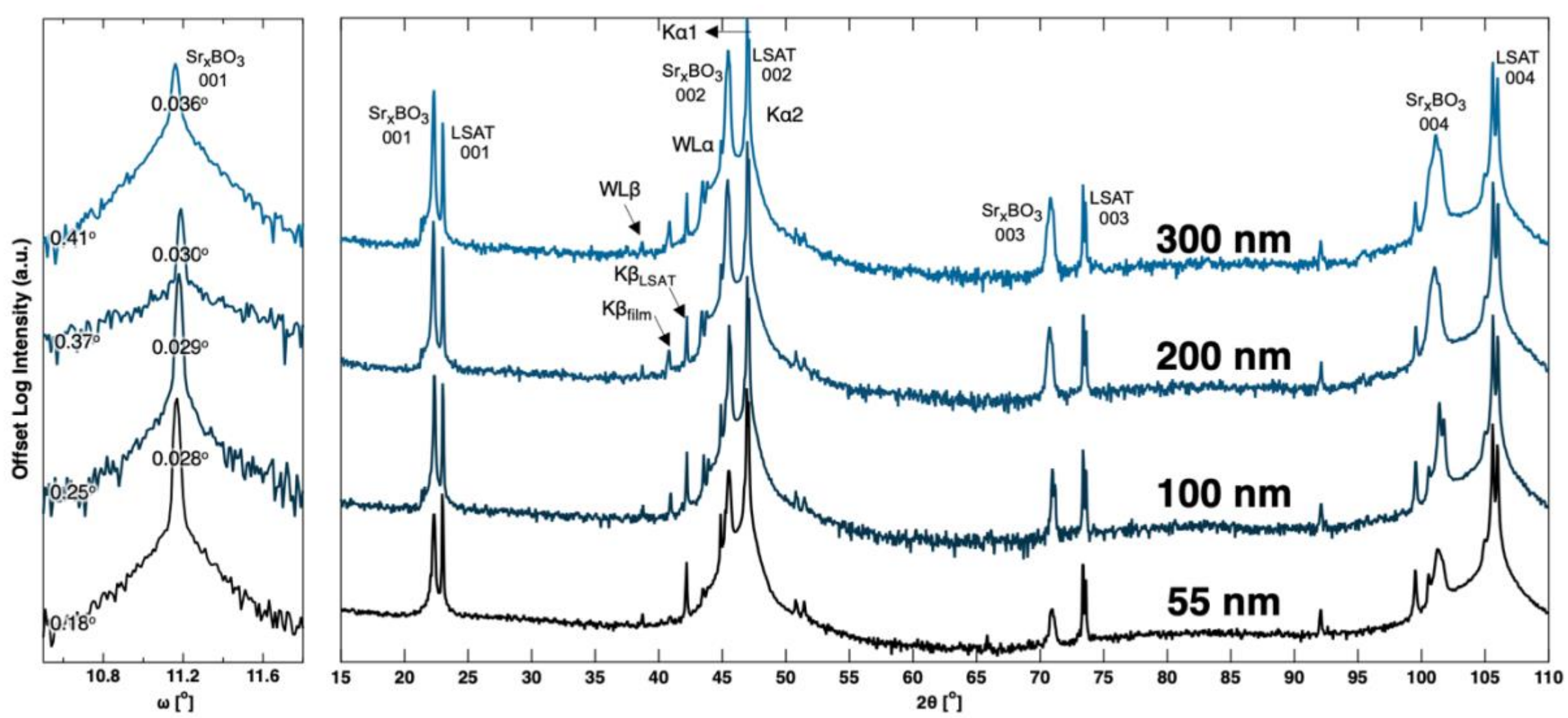

**Figure S10.** BBHD X-ray diffraction patterns and rocking omega curves for a series of $Sr_{0.95}BO_3$ thin films grown thick up to 300 nm

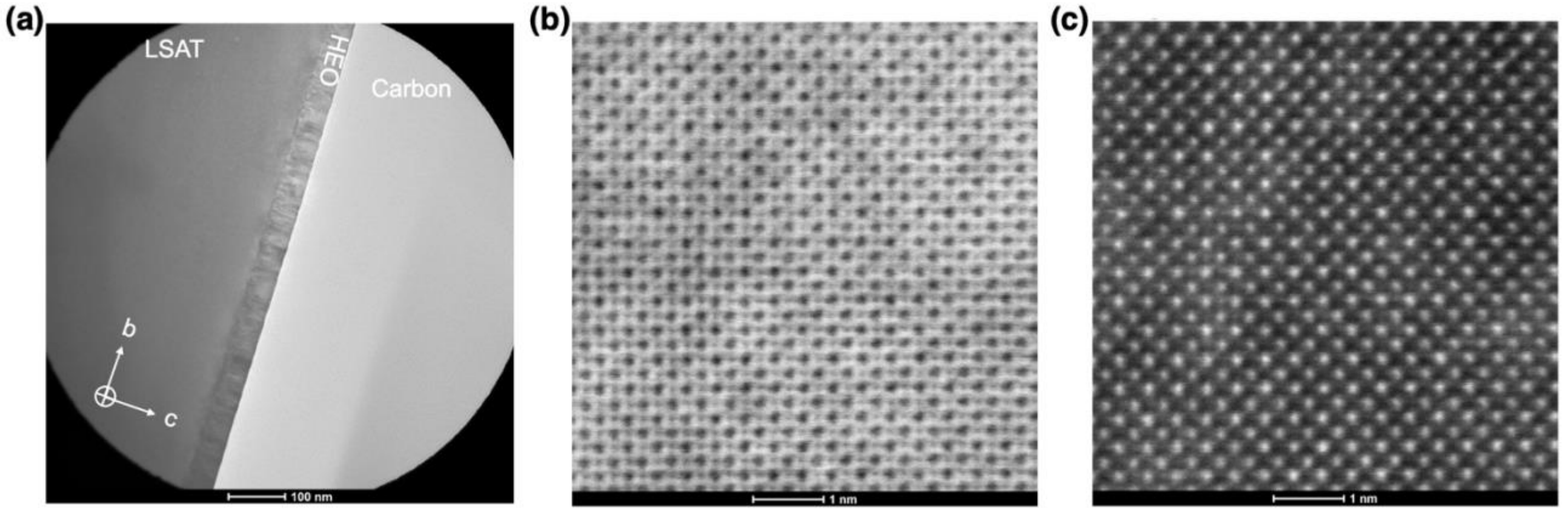

**Figure S11.** (a) the selected area corresponding to the electron diffraction in Figure 2d, (b) an Annular Bright Field STEM image that highlights the positions of both oxygen atoms and cations, while (c) shows the simultaneously acquired HAADF STEM image.

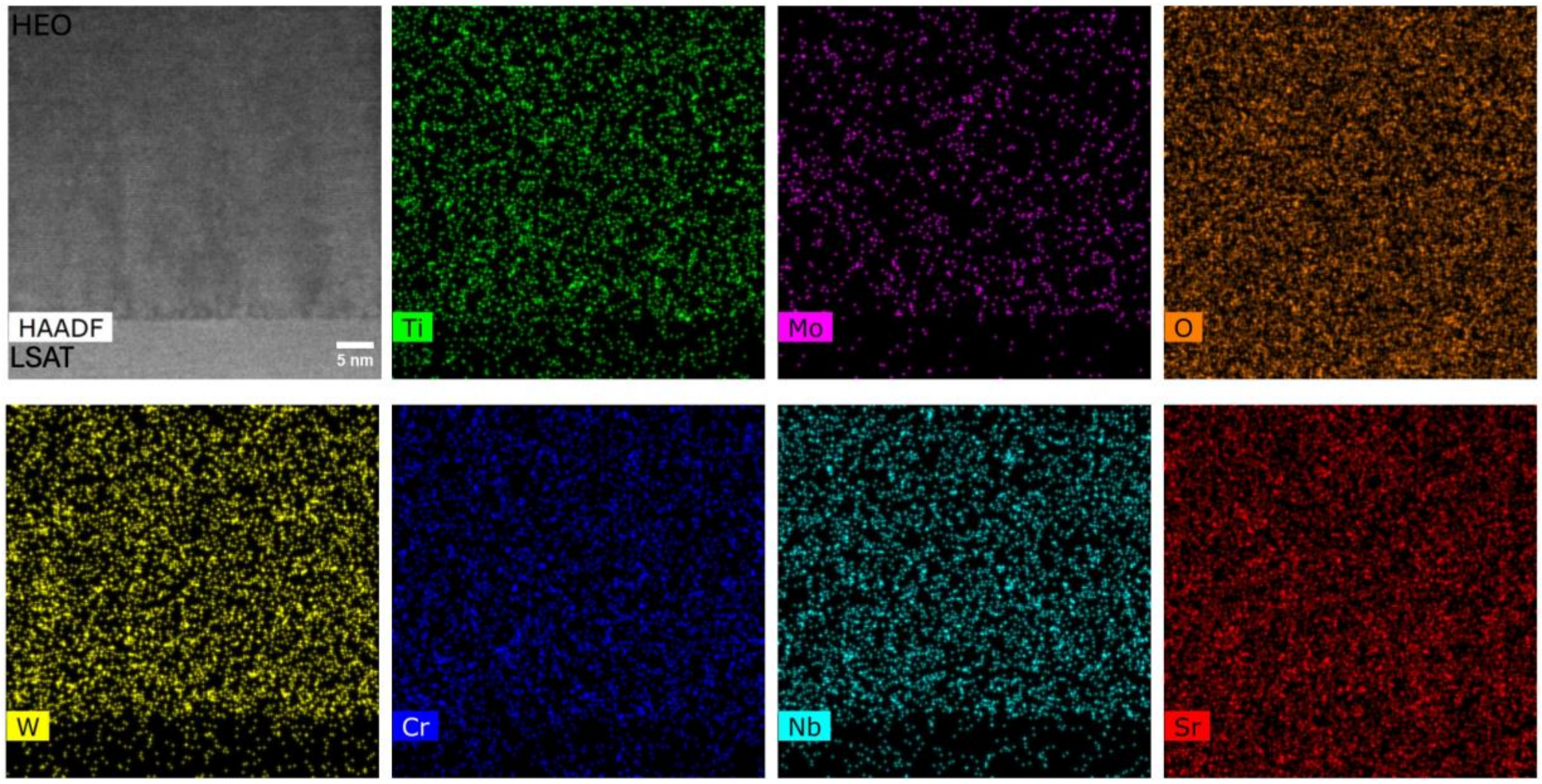

**Figure S12.** EDX maps of all individual elements in $Sr_{0.95}BO_3$ and the corresponding HAADF image

Note 7: Second harmonic generation and "global" centrosymmetry

We conducted an optical second harmonic generation (SHG) experiment using an 800 nm fundamental laser (Spectra-Physics Ti: sapphire gain medium, 80 fs, 1 kHz) at 45° oblique incidence reflection geometry. The setup is shown in Figure S13. The laser was focused with a 10 cm convex lens, achieving a beam diameter of 25 μm at the focus, measured using a knife-edge technique. The laser-induced surface damage threshold was determined to be 225 GW/cm² by exposing the thin film to various fluences and inspecting for damage under an optical microscope. Nonlinear second harmonic intensity was then measured as a function of the incident intensity.

SHG is a nonlinear optical process where a material generates polarization at frequency 2ω from an incident ω beam, described by second-order optical susceptibility, $P_i^{2\omega} = d_{ijk}E_j^{\omega}E_k^{\omega}$.[19] In a non-centrosymmetric material, the intensity dependence of the second harmonic should be quadratic relative to the incident beam ($I^{2\omega} = \mathrm{A}^{*}(I^{\omega})^2$). Measurements from this experiment showed that the generated $2\omega$ beam intensity did not follow a quadratic dependence expected of a non-centrosymmetric material (Figure S13), indicating that $Sr_{0.95}BO_3$ films on LSAT are centrosymmetric within the 25 $\mu$m probe region.

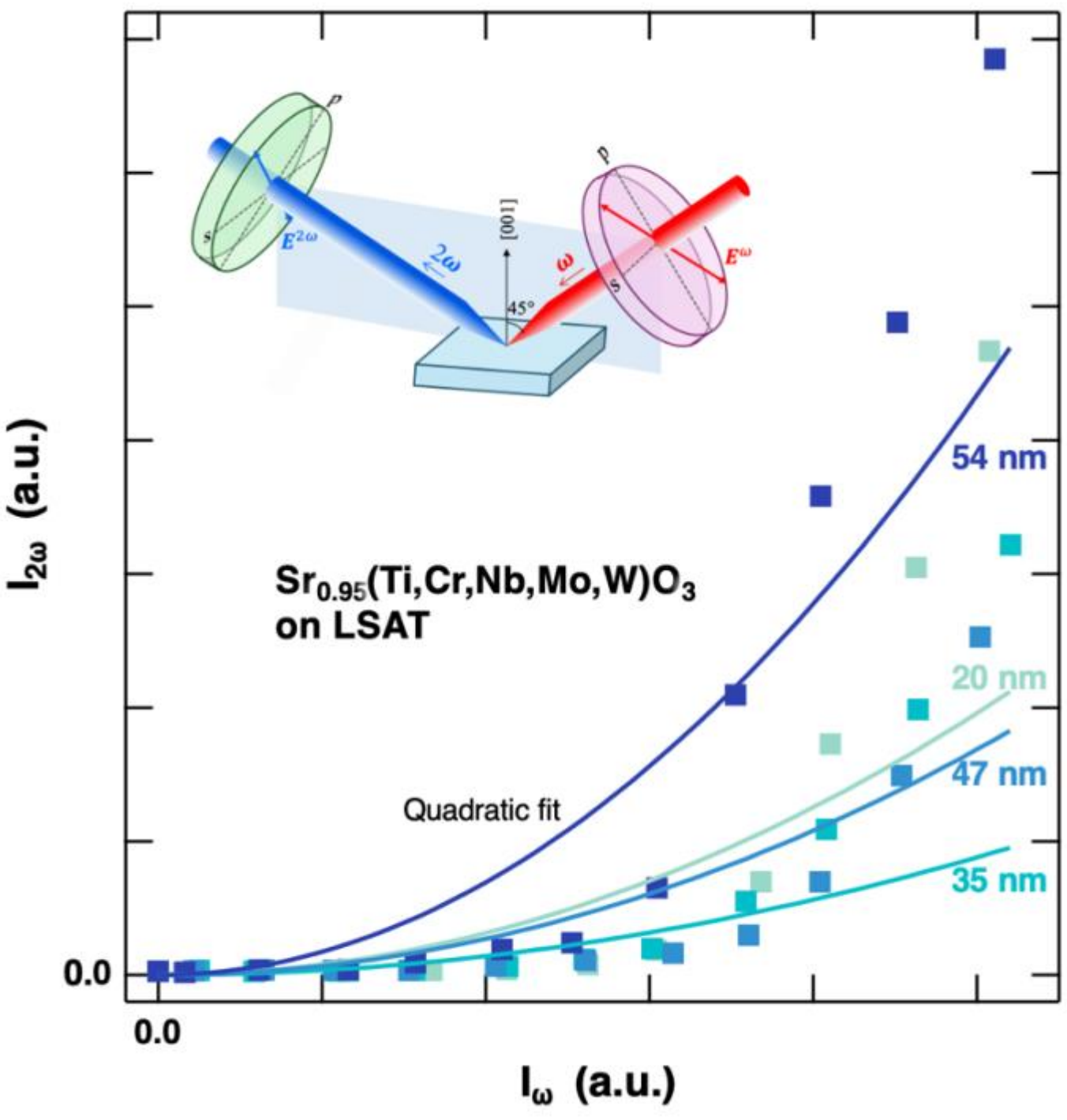

**Figure S13.** SHG measurements showing the intensity of generated second harmonic response at frequency 2ω as a function of incident laser intensity at the fundamental frequency ω for $Sr_{0.95}BO_3$ films on LSAT shown in Figure 2.

## Note 8: $Sr_{0.95}BO_3$ XPS data analysis

Figure S15 depicts the XPS spectra for Sr 3d, Ti 2p, and Cr 2p3/2. The spectra confirm that Sr predominantly adopts the 2+ oxidation state, Ti primarily exists in the 4+ oxidation state, and Cr is mainly present in the 3+ oxidation state. A summary of all observed oxidation states from XPS analysis is provided in Table S1. The fitting process for Nb, Mo, and W is more complex, with the final results of the fitting detailed in the main text (**Figure 3**). We performed the Nb fitting using $SrNbO_3$ as a reference, as described in Note 10. For Mo and W, we utilized an available set of standard reference data alongside line shapes we derived from fitting of $MoO_3$, $MoO_2$, $MoS_2$, $SrMoO_3$ and $WO_3$ to fit the $Sr_{0.95}BO_3$ peaks accurately. To ensure reliability, XPS data fitting was tested for self-consistency across 30 $Sr_{0.95}BO_3$ samples. Additionally, the entire valence determination experiment was validated by applying charge neutrality, as detailed in **Table S2**, using the same sample analyzed in the EDS mapping. The final result demonstrates that the summation of positive and negative charges—accounting for all measured valence states and the intentional Sr vacancy concentration—is equal and opposite, satisfying charge neutrality.

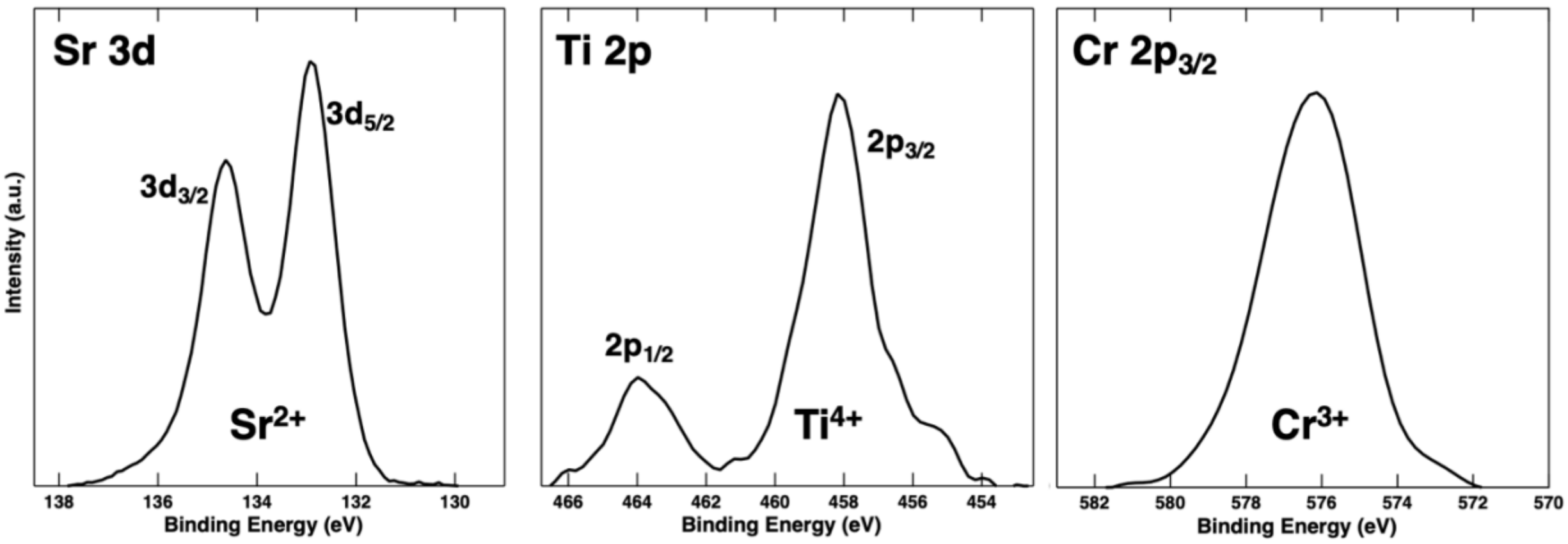

**Figure S14.** XPS spectra of Sr, Ti and Cr in 35 nm $Sr_{0.95}BO_3$

Table S1: $Sr_{0.95}$(Ti,Cr,Nb,Mo,W)$O_3$ cation oxidation states observed in XPS

| Cation | Oxidation states observed in XPS |
|---|---|
| **Sr** | 2+ |
| **Ti** | 4+ |
| **Cr** | 3+ |
| **Nb** | 2+, 4+ and 5+ (possibly 3+) |
| **Mo** | 4+, 5+ and 6+ |
| **W** | 4+ and 6+ |

Table S2: $Sr_x$(Ti,Cr,Nb,Mo,W)$O_{3+d}$ charge neutrality calculations

| element | stoichiometry | average oxidation per element | total valence |
|---|---|---|---|
| **Sr** | 0.92 | 2 | 1.84 |
| **Ti** | 0.18 | 4 | 0.72 |
| **Cr** | 0.19 | 3 | 0.57 |
| **Nb** | 0.21 | 3.97 | 0.8337 |
| **Mo** | 0.2 | 4.94 | 0.988 |
| **W** | 0.21 | 5.54 | 1.1634 |
| **O** | 3.058 | -2 | -6.116 |
| | | **sum** | -0.0009 |

Note 9: $SrNbO_3$ growth and optical properties from $Sr_xNbO_3$ targets

Similar to Note 4, we grew $Sr_xNbO_3$ thin film series using PLD from targets prepared with x = 0.5, 0.75 and 1. Films grown from $Sr_{0.75}NbO_3$ target, exhibit the highest crystalline quality (Figure S16) and best optical performance while maintaining low extinction coefficient k (Figure S17) and high electrical conductivity of 300-400 µOhm.cm at room temperature. Therefore, we choose the film grown from $Sr_{0.75}NbO_3$ target in Figure S16 for the comparison with the high entropy system in Figure 4.

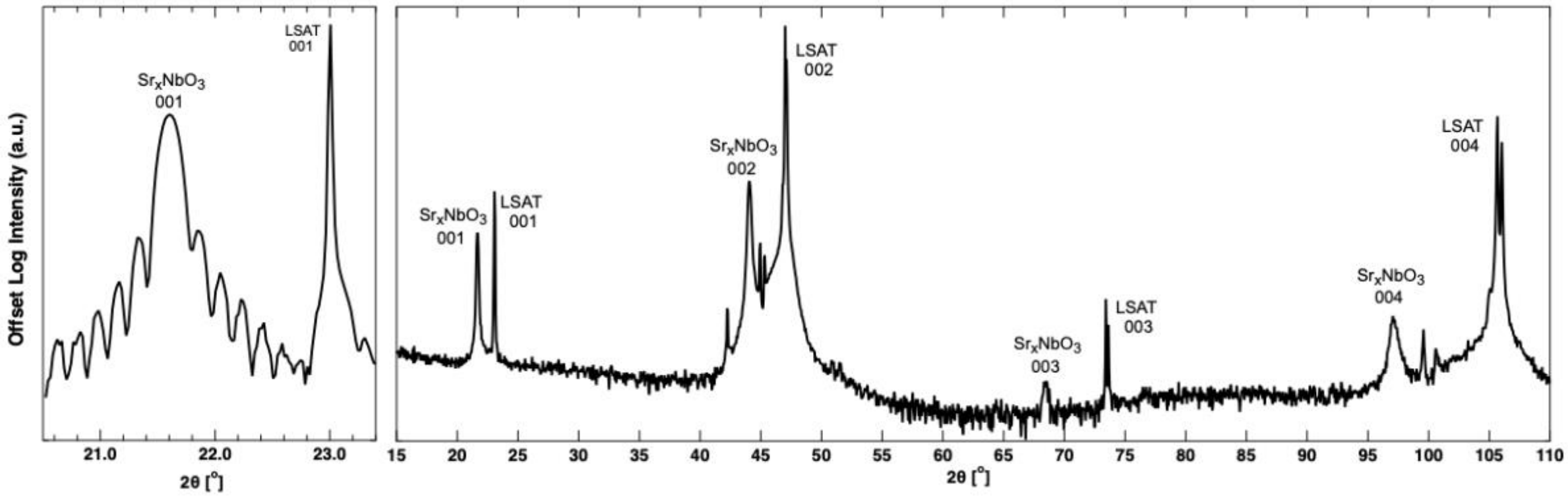

**Figure S15.** BBHD and high-resolution X-ray diffraction scan of $Sr_xNbO_3$ film grown on LSAT from $Sr_{0.75}NbO_3$ target

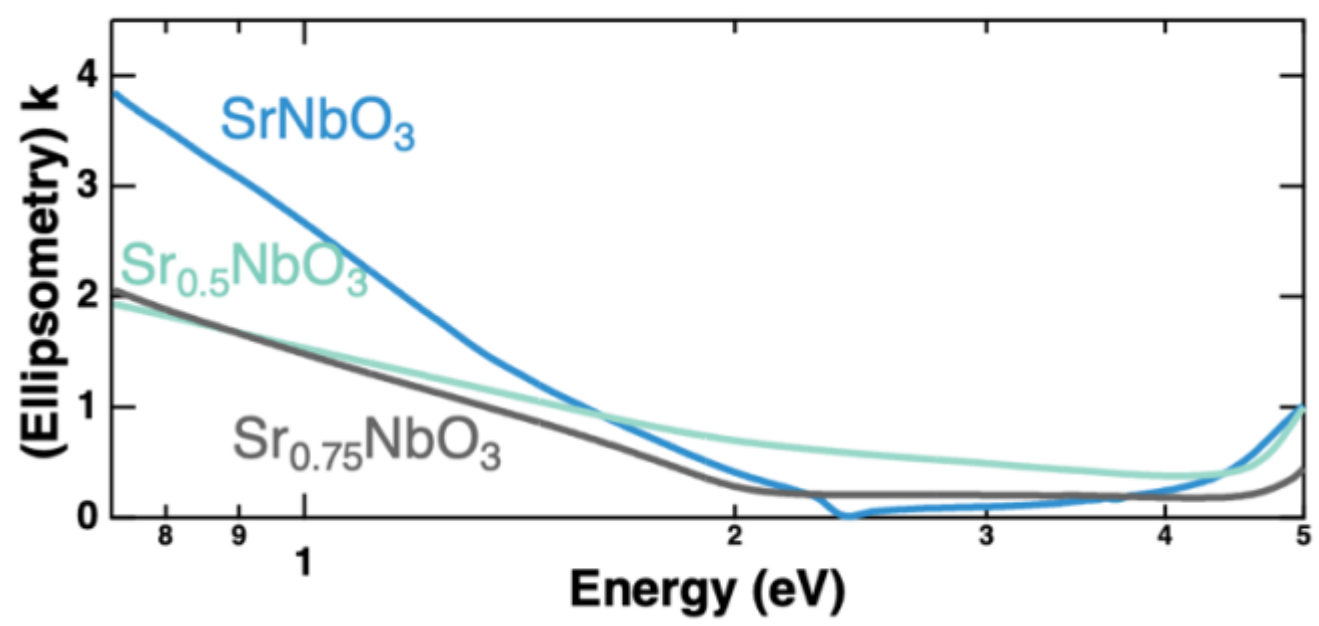

**Figure S16.** Extinction coefficient of three films grown from $SrNbO_3$, $Sr_{0.75}NbO_3$ and $Sr_{0.5}NbO_3$ ceramic targets, respectively.

## Note 10: $Nb^{2+}$ evolution in $Sr_xNbO_3$ X-ray photoelectron spectroscopy

To achieve a proper fit of the Nb 3d spectra, we followed the approach outlined by Roth et al.[18], incorporating contributions from $Nb^{3+}$ and $Nb^{2+}$ states. As shown in Figure S18, the $Nb^{2+}$ concentration increases with rising Sr vacancies on the A-site in $Sr_xNbO_3$. We applied the same fitting model as a starting point for Nb in $Sr_{0.95}BO_3$. However, for the high-entropy oxide samples, adding $Nb^{3+}$ did not significantly enhance the fit, so we excluded it to avoid overfitting.

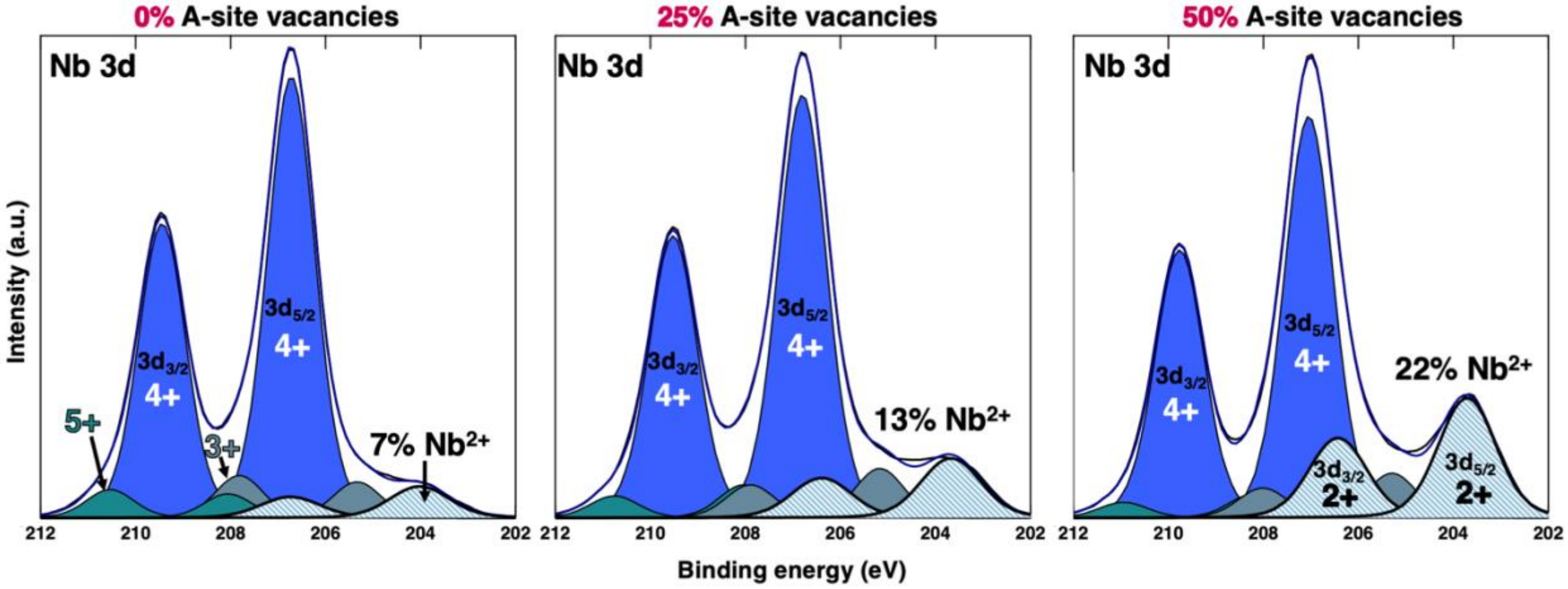

**Figure S17.** XPS fittings for the Nb 3d spectra of three films with increasing Sr vacancies, demonstrate a corresponding increase in $Nb^{2+}$ concentration

Note 11: X-ray absorption near edge structure (XANES)

XANES measurements were conducted using an easyXAFS300+ spectrometer (Renton, WA), operating in fluorescence mode with a Mo X-ray tube set at 40 kV and 30 mA. A minimum of 75 scans were performed for the Sr K-edge (16,105 eV) and the W $L_3$ -edge (10,207 eV). Data analysis was carried out with Athena software from the Demeter package[20], while peak fitting was performed using Larch[21]. Ta metal (9,881 eV) was employed for energy calibration of the W $L_3$ -edge. The XANES results are illustrated in Figure S14. The Sr K-edge, in Figure S14(a), corresponds to Sr in a high coordination environment, closely resembling strontium carbonate or doped aragonite, where Sr exhibits 9-fold coordination[22]. Figure 14(b) shows the W $L_3$ absorption edge, which corresponds to an electronic transition from the $2p_{3/2}$ to an empty 5d state. The splitting of the d states is observed in the second derivative of the W $L_3$ spectra (Figure 2). The energy difference between the two minima in this derivative, measured to be 3.7 eV, indicates a distorted octahedral environment for W, based on comparisons with similar systems[23]. Further analysis involved fitting the W $L_3$ white line with two Lorentzian functions centered at the energy values of the minima, as shown in Figure S14(b). The areas under these Lorentzian functions reflect the absorption intensities of the $t_{2g}$ and $e_g$ orbitals. Compared to other materials with distorted W octahedral environments, the absorption intensity of the $t_{2g}$ orbital is lower than expected. Additionally, the $E_0$ shift of the W $L_3$ absorption edge was used to estimate the average oxidation state. The $E_0$ shift, calculated as 6.4 eV, suggests that the majority oxidation state of W in this sample is 6+.

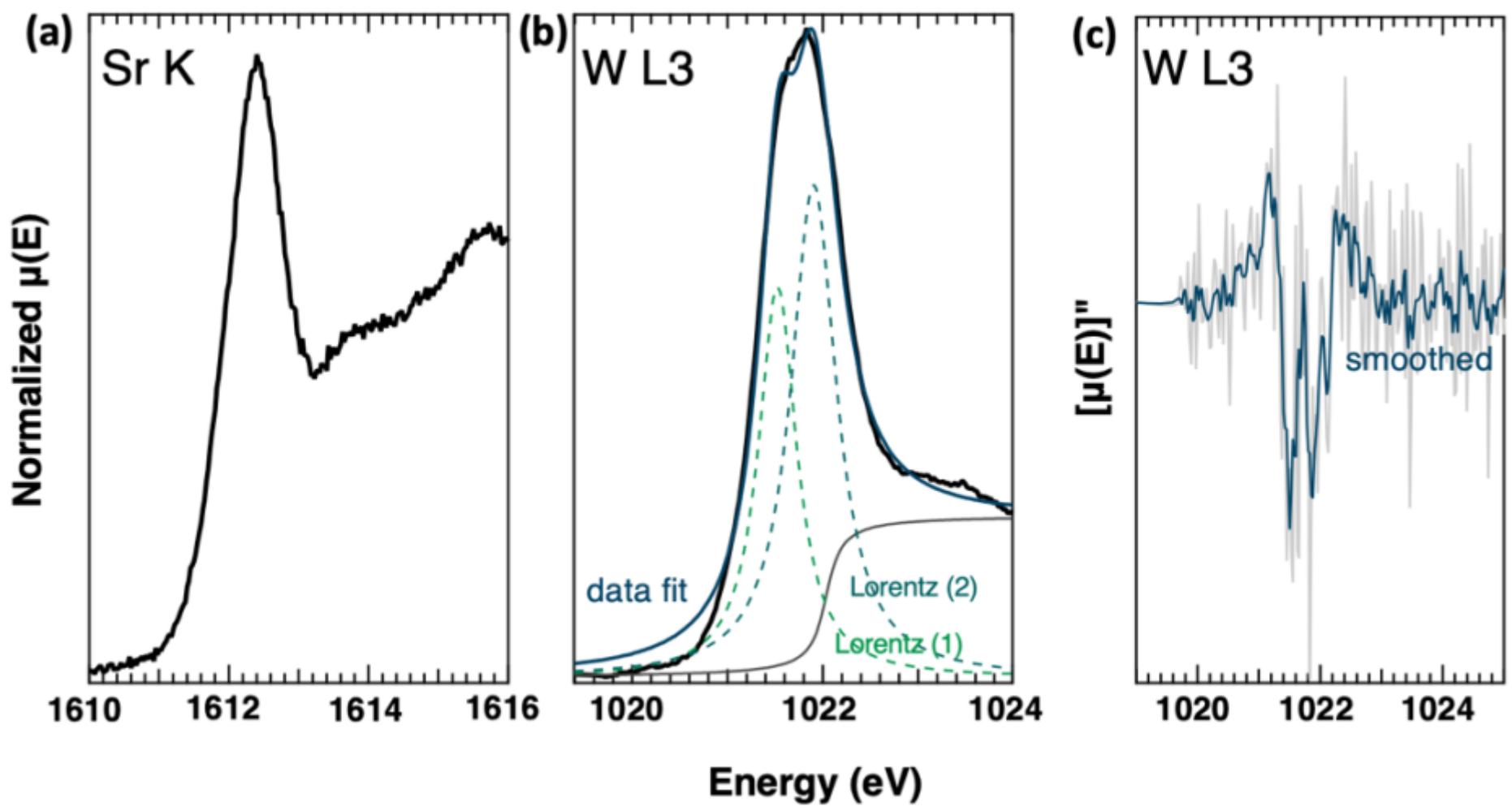

**Figure S18.** XANES Sr K absorption edge (a) and WL3 absorption edge (b). The second derivative of WL3 is plotted in (c).

Note 12: Optical coefficient for $Sr_{0.95}BO_3$ films on LSAT

The real part of the dielectric function (ε1) in Figure S19(a) indicates a screened plasma frequency for the high-entropy films well below the visible range, around 1.33 eV, which is comparable to $SrVO_3$ and red-shifted compared to $Sr_xNbO_3$. The imaginary part (ε2) and the extinction coefficient (k) in Figure S19(b) and (c) suggest very low absorption in the visible and UV regimes, with a sharp increase at higher photon energies attributed to $E_{O2p\text{-}t2g}$ interband transitions. Figure S19 (d) shows the extinction coefficient (k), calculated from ultraviolet-visible (UV-Vis) spectroscopy in transmission mode. The extinction coefficient (k) obtained from both ellipsometry in reflection mode and UV-Vis in transmission mode shows remarkable agreement, strengthening our confidence in the measurements and fittings. $Sr_{0.95}BO_3$ samples exhibit slightly lower k values compared to $Sr_xNbO_3$ particularly at lower energies, suggesting higher optical transmission in the high-entropy samples.

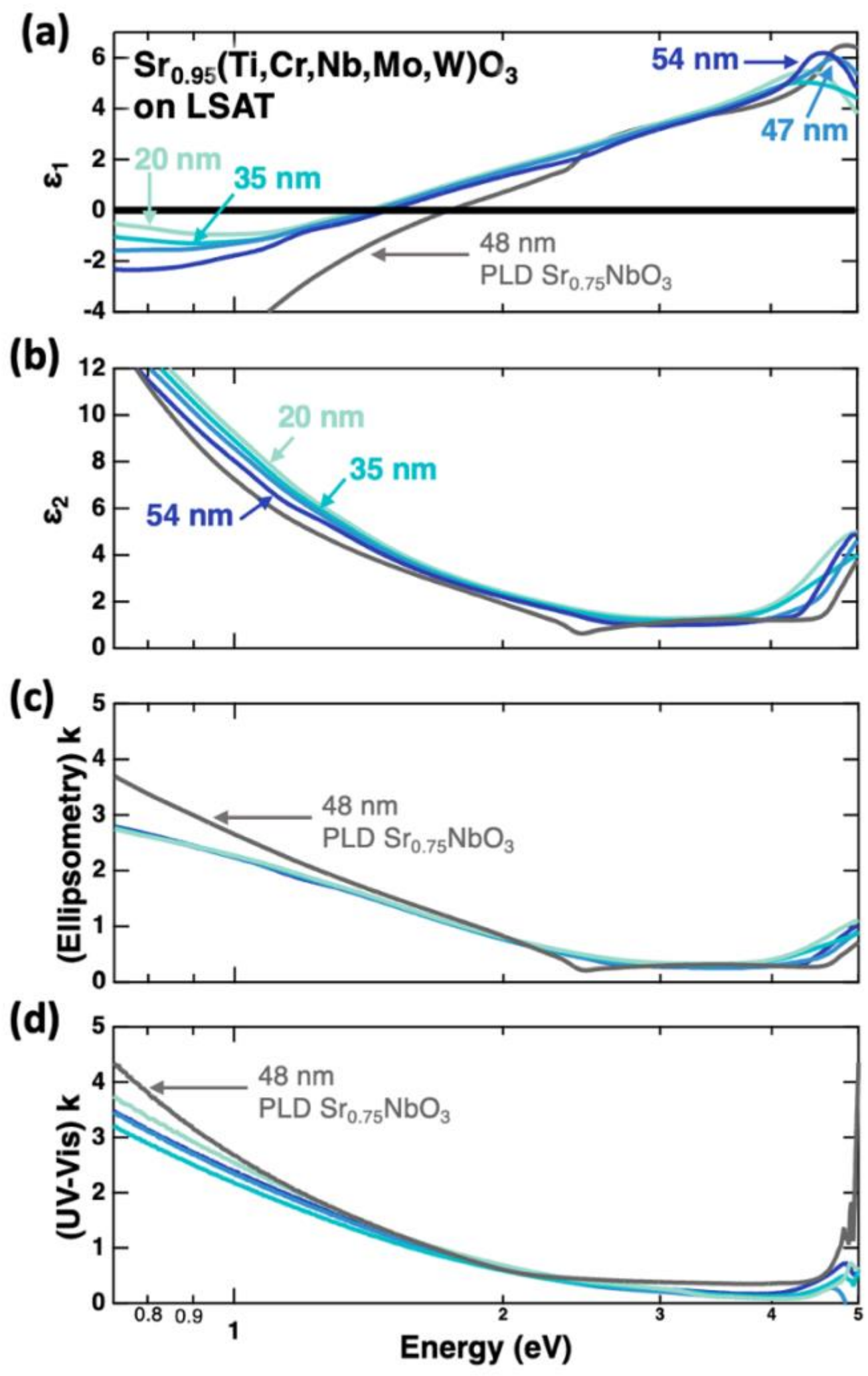

**Figure S19.** (a)-(c) summarizes the ellipsometry optical properties of $Sr_{0.95}BO_3$ in comparison to PLD grown $Sr_xNbO_3$ (this study) and (d) shows the extinction coefficient calculated from UV-Vis.

Note 13:

Everything matters: $Sr_xBO_3$ resistivity distribution over three years and 200 samples

Due to the high metastability of $Sr_xBO_3$, the thin film growth conditions are highly sensitive and have a significant impact on the resulting electronic properties of the samples. For example, given the oxygen-rich nature of the targets and the tendency of Mo and W to favor the $ABO_4$ structure, it is critical to maintain an optimal flow of reducing gas (in this case, Ar) to promote the formation of the $ABO_3$ phase. However, even a slight excess in gas flow could lead to the reduction of W and Mo. Similar considerations apply to parameters such as growth rate, time on the heater, growth temperature, and laser fluence.

Figure S19 shows a histogram illustrating the resistivity distribution of 200 samples grown under varied conditions. Films with room-temperature resistivity below 1000 μΩ·cm were produced under specific conditions: a substrate temperature of 850°C measured by thermocouple and 700°C measured by external pyrometer, an Ar flow of 40 sccm and pressure of 50 mTorr (the vacuum chamber background pressure is ~8 x $10^{-8}$ Torr), laser fluence of 1.4 J/cm², and a growth rate of 22 nm/min. Post-growth, the films were quenched to ambient conditions within six minutes after the final laser pulse to minimize any post-synthesis annealing effects.

| **Growth Parameter** | **Range Explored** |
| --- | --- |
| Substrate temperature (thermocouple) | 650 – 950 °C |
| Laser fluence | 0.8 – 1.5 J/cm² |
| Growth rate | 7 – 30 nm/min |
| Ar gas flow | 0 – 50 sccm |
| Ar gas pressure | 1 – 100 mTorr |
| Total time in the chamber | 2 hrs – 12 min |

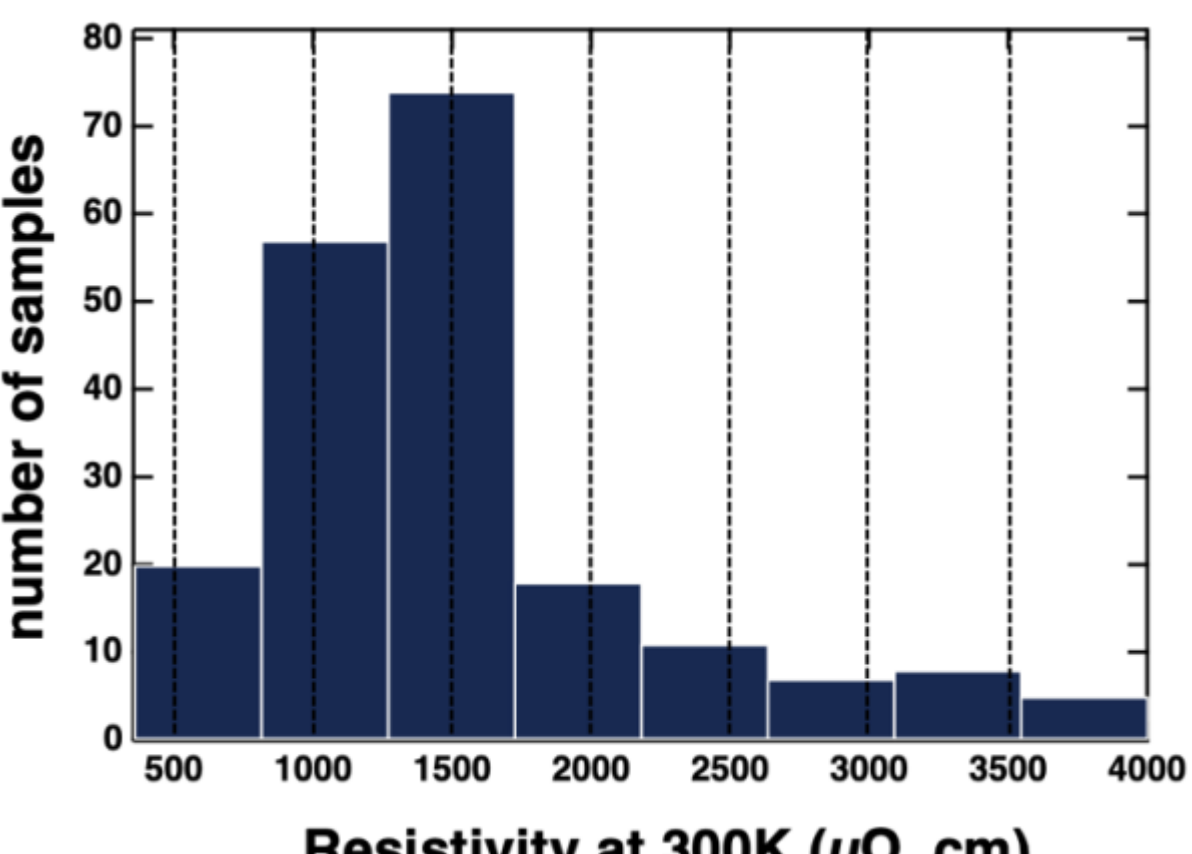

**Figure S20.** $Sr_xBO_3$ room temperature resistivity histogram over three years and 200 samples

Note 14:

Short-range chemical ordering in $SrBO_3$

**Methods**

Density Functional Theory (DFT) calculations for the cluster expansion were carried out using version 6.4.2 of the Vienna Ab initio Simulation Package VASP software[13], within the spin-polarized generalized gradient approximation (GGA) and the Perdew-Burke-Ernzerhof (PBE) parametrization.[24] Hubbard U corrections were applied to the 3d-orbitals of Cr (U=3.7 eV), Mo (U=4.38 eV), and W (U=6.2 eV), and energies were calculated following the methodology employed by the Materials Project (database release v2021.11.10).[25,26] Ionic cores were represented using Projector Augmented Wave (PAW) pseudopotentials[13,27,28], and the energy cutoff for the plane-wave basis set was 680 eV. The Brillouin zone was sampled using a Monkhorst-Pack grid with a k-spacing of 0.22 $Å^{-1}$. Atomic configurations were relaxed at zero pressure, using a maximum force threshold of 10 meV·$Å^{-1}$, while allowing ionic positions, cell shape, and cell volume to relax. To minimize the Pulay stress, relaxation runs were repeated until the volume change between two consecutive runs was less than 1%. Finally, the enthalpy of mixing per formula unit was computed as

$$\Delta H_{\mathrm{mix}}^{\mathrm{DFT}} = \frac{E_{conf}^{DFT} - \sum_M N_M E_{\mathrm{SrMO_3}}^{DFT}}{\sum_M N_M} \quad (1),$$

where $E_{conf}^{DFT}$ and $E_{\mathrm{SrMO_3}}^{DFT}$ are energies of the perovskite compounds calculated using DFT, and $N_M$ is the number of atoms of type $M$ in the perovskite configuration denoted "conf".

The cluster expansion (CE) method was employed to evaluate the enthalpies of mixing of configurations containing thousands of atoms.

This method uses an orthogonal basis of cluster functions to describe cluster interactions in a crystal structure where $\gamma_{\alpha_i}(\sigma_i)$ is an orthogonal basis function of a cluster, $\boldsymbol{\alpha}$, within a specific configuration, $\boldsymbol{\sigma}$, and the product is calculated over all the lattice sites. The mixing energies are then modeled as a linear combination of the cluster functions (eqn. 2), where $J_{\boldsymbol{\alpha}}$ are the cluster interactions:

$\Gamma_{\alpha}(\boldsymbol{\sigma}) = \prod_i \gamma_{\alpha_i}(\sigma_i)$ (2) and $\Delta H_{mix}^{CE}(\boldsymbol{\sigma}) = \sum_{\boldsymbol{\alpha}} J_{\boldsymbol{\alpha}} \Gamma_{\boldsymbol{\alpha}}(\boldsymbol{\sigma})$ (2).

The effective cluster interactions were extracted using Automatic Relevance Determination Regression (ARDR), from a database of 1478 atomic configurations relaxed within DFT. Starting from the orthorhombic unit cell of the Pnma space group, containing 4 formula units, symmetrically unique configurations over the composition space $\mathrm{Sr}Ti_{x_{Ti}}Cr_{x_{Cr}}Nb_{x_{Nb}}Mo_{x_{Mo}}W_{x_W}O_3$ ($\sum_M x_M = 1$) were generated using the structure enumeration module[29,30] within the Integrated Cluster Expansion Toolkit Python package.[31] We used orthorhombic unit cells to avoid inconsistencies between energies evaluated in small and large cells with cubic symmetry. All 175 configurations containing 4 formula units were included. Among structures made of 8 formula units, 1303 binary $\mathrm{Sr}X_{x_X}Y_{x_Y}O_3$ and quinary $SrTi_{x_{Ti}}Cr_{x_{Cr}}Nb_{x_{Nb}}Mo_{x_{Mo}}W_{x_W}O_3$ configurations were randomly sampled and relaxed. The CE model was built using all pair and triplet clusters such that the largest distance between two atoms in a cluster is less than 8.5 Å for pairs (140 pairs) and 6.5 Å for triplets (640 triplets). Cluster model predictions are reported in Figure S21(a). They are in reasonable agreement with the DFT calculations. Using the fitted cluster expansion model, 5 independent $Sr(Ti,Cr,Nb,Mo,W)_{1/5}O_3$ atomic configurations, each containing 5000 atoms, were equilibrated at 1500 K, 2000K, and 3000K using Metropolis Monte Carlo simulations within the canonical ensemble.[32]

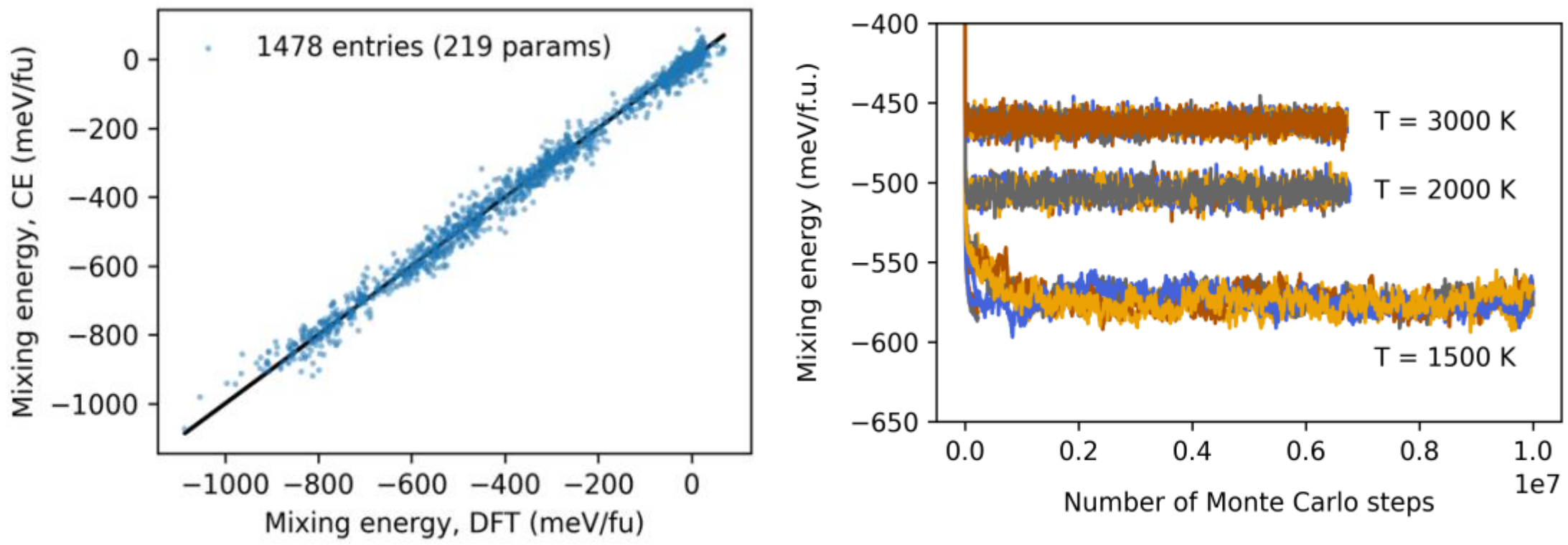

**Figure S21.** (a) Parity plot comparing the DFT and CE energies of the 1478 configurations used to fit the CE model (made of 219 non-zero parameters after ARDR). The out-of-sample Root Mean Square Error is equal to 6.5 meV/atom and the coefficient of determination, $R^2$, is 0.99, and (b) Evolution of mixing energy as a function of Monte Carlo step number for the equimolar $Sr(Ti,Cr,Nb,Mo,W)_{1/5}O_3$ configurations. 5 independent samples, with an initially random distribution of transition metal elements, were equilibrated at 1500 K, 2000 K, and 3000 K.

## Comment on equilibration temperature

We employ the Warren-Cowley parameter[33], $w_{\alpha\beta,n}$, to quantify chemical ordering, defined as

$$w_{\alpha\beta,n} = 1 - \frac{p_{\alpha\beta,n}}{c_\beta} \quad (3)$$

where $p_{\alpha\beta,n}$ is the probability of finding a type $\beta$ atom with $c_\beta$ concentration in the $n^{\text{th}}$ nearest neighbor shell of a type $\alpha$ atom. In the absence of chemical ordering, $p_{\alpha\beta,n} = c_\beta$ yielding $w_{\alpha\beta,n} = 0$. Therefore, nonzero $w_{\alpha\beta,n}$ suggests chemical ordering, with negative values indicating clustering tendencies and positive values suggesting repulsive interactions between species $\alpha$ and $\beta$.

Although the films are not equilibrated at a controlled temperature during deposition, they grow as crystalline solids. This suggests that they undergo some degree of energy relaxation via thermally activated processes, such as vacancy diffusion, and supports assessing tendencies for chemical ordering in the out-of-equilibrium films using thermally equilibrated atomistic configurations. Additionally, temperature modulates the amplitude of ordering, but preserves the overall trends, as demonstrated by the Warren-Cowley analyses at different temperatures reported in Figure S22. Therefore, the discussion of ordering does not depend on the choice of temperature.

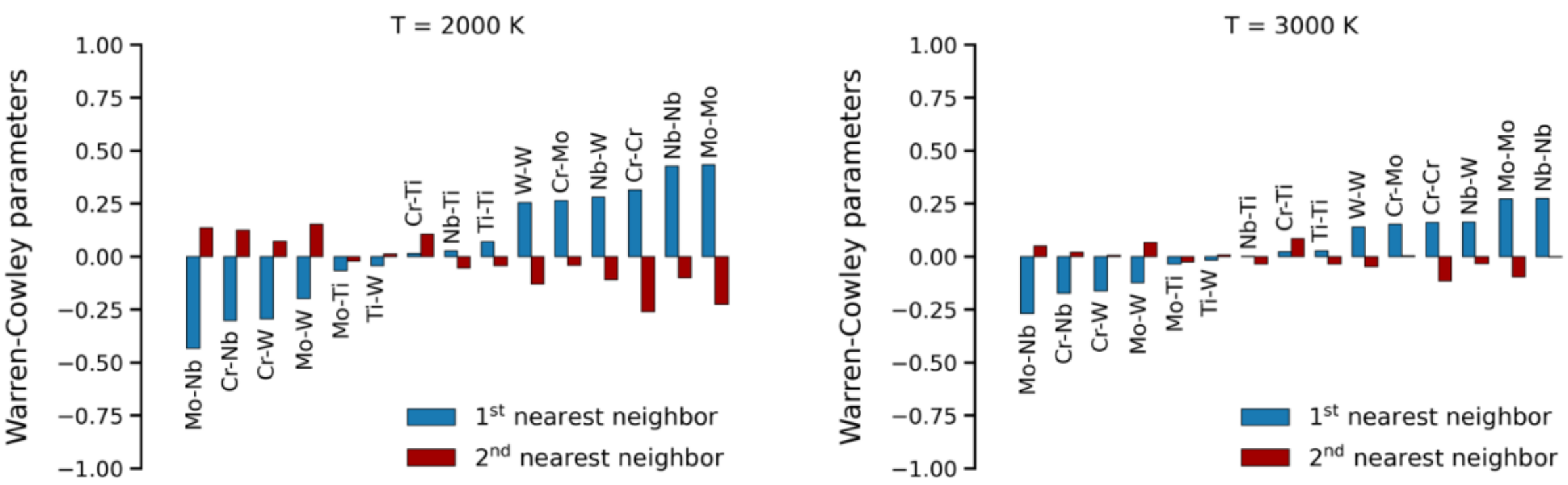

**Figure S22.** Warren-Cowley parameters for pairs of species in the first and second nearest neighbor shells. As temperature increases, the amplitudes of Warren-Cowley parameters decrease. On the contrary, their ordering, from most attractive to most repulsive pairs, is almost unaltered.